\documentclass[aps,twocolumn,superscriptaddress]{revtex4-1}

\usepackage{bm,mathrsfs}
\usepackage{graphicx}
\usepackage{epsfig}
\usepackage{amsmath,bbm}
\usepackage{amsfonts,amssymb}
\usepackage{times}
\usepackage{dsfont}
\usepackage{enumitem}  
\usepackage[colorlinks,linkcolor=blue,citecolor=blue]{hyperref}
\usepackage{comment}

\newcommand{\eqn}[1]{
\begin{eqnarray}
	#1
\end{eqnarray}
}

\begin{document}

\title{Variational principle for quantum impurity systems in and out of equilibrium: application to Kondo problems}

\author{Yuto Ashida}
\email{ashida@cat.phys.s.u-tokyo.ac.jp}
\affiliation{Department of Physics, University of Tokyo, 7-3-1 Hongo, Bunkyo-ku, Tokyo 113-0033, Japan}
\author{Tao Shi} 
\email{tshi@itp.ac.cn}
\affiliation{CAS Key Laboratory of Theoretical Physics, Institute of Theoretical Physics, Chinese Academy of Sciences, P.O. Box 2735, Beijing 100190, China}
\affiliation{Max-Planck-Institut f\"ur Quantenoptik, Hans-Kopfermann-Strasse. 1, 85748 Garching, Germany}
\author{Mari Carmen Ba\~nuls}
\affiliation{Max-Planck-Institut f\"ur Quantenoptik, Hans-Kopfermann-Strasse. 1, 85748 Garching, Germany}
\author{J. Ignacio Cirac}
\affiliation{Max-Planck-Institut f\"ur Quantenoptik, Hans-Kopfermann-Strasse. 1, 85748 Garching, Germany}
\author{Eugene Demler}
\affiliation{Department of Physics, Harvard University, Cambridge, Massachusetts 02138, USA}

\begin{abstract}
We provide a detailed formulation of the recently proposed variational approach [Y.~Ashida et al., Phys.~Rev.~Lett.~121,~026805~(2018)] to study ground-state properties and out-of-equilibrium dynamics for generic quantum spin-impurity systems. Motivated by the original ideas by Tomonaga, Lee, Low, and Pines, we construct a canonical transformation that completely decouples the impurity from the bath degrees of freedom. By combining this transformation with a Gaussian ansatz for the fermionic bath, we obtain a family of variational many-body states that can efficiently encode the strong entanglement between the impurity and fermions of the bath. We give a detailed derivation of equations of motions in the imaginary- and real-time evolutions on the variational manifold. We benchmark our approach by applying it to investigate ground-state and dynamical properties of the anisotropic Kondo model and compare results with those obtained using matrix-product state (MPS) ansatz. We show that our approach can achieve an accuracy comparable to MPS-based methods with several orders of magnitude fewer variational parameters than the corresponding MPS ansatz. Comparisons to the Yosida ansatz and the exact solution from the Bethe ansatz are also discussed.  We use our approach to investigate the two-lead Kondo model and analyze its long-time spatiotemporal behavior and  the conductance behavior at finite bias and magnetic fields. The obtained results are consistent with the previous findings in the Anderson model and the exact solutions at the Toulouse point.
\end{abstract}

\maketitle

\section{Introduction}
Out-of-equilibrium phenomena in quantum many-body systems are an active area of research in both  ultracold gases  \cite{EM11,CM12,FuT15,KAM16,RL17} and traditional solid-state physics  \cite{DFS02,THE11,LC11,IZ15,MMD17}. A broad class of problems that correspond to a quantum impurity coupled to the many-body environment are particularly important and have been at the forefront of condesned matter physics starting with the pioneering paper by Kondo \cite{KJ64}.
They have proven crucial to the understanding of thermodynamic properties in strongly correlated materials   \cite{AK75,HAC97,LH07,GP08,SQ10}, transport phenomena  \cite{GL88,NTK88,MY93,LW02,YLH04,GGD98,CSM98,SF99,WWG00,RMP07,KAV11,KAV12}  and decoherence  \cite{LAJ87,LD98,WZ07}  in nanodevices, and lie at the heart of formulating dynamical mean-field theory \cite{GA96} (DMFT).

\begin{figure}[b]
\includegraphics[width=50mm]{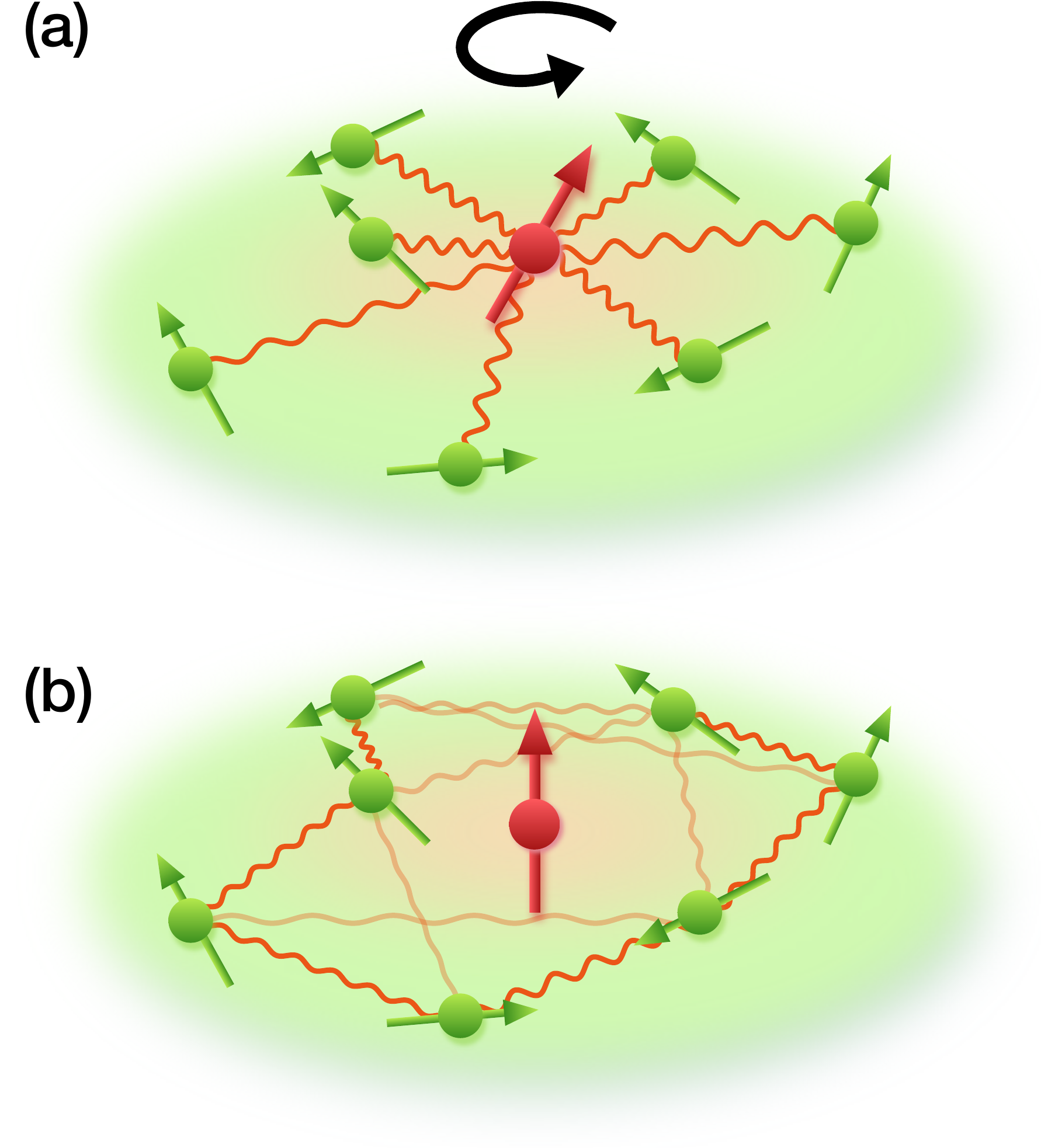} 
\caption{\label{fig_schem1}
Schematic illustration of the canonical transformation introduced in our variational approach. (a) In the original frame, the localized impurity spin can interact with mobile bath particles in an arbitrary manner, generating strong entanglement between the impurity spin and bath. (b) After employing the canonical transformation, we can move to the ``corotating" frame of the impurity in which the impurity dynamics can be made frozen at the expense of introducing an interaction among bath particles.
}
\end{figure}

A variational approach is one of the most successful and powerful approaches for solving many-body problems. Its guiding principle is to design a family of variational states that can efficiently capture essential physics of the quantum many-body system while avoiding the exponential complexity of the exact wavefunction. In quantum impurity problems, such variational studies date back to Tomonaga's treatment \cite{TS47} of the coupling between mesons and a single nucleon. Soon after that, Lee, Low and Pines \cite{LLP53} (LLP) applied similar approach to the problem of a polaron, a mobile spinless impurity interacting with phonons. Their key idea is to take advantage of the total-momentum conservation by transforming to the comoving frame of the impurity via the unitary transformation 
\eqn{\label{LLP}
\hat{U}_{\rm LLP}=e^{-i\hat{\bf x}\cdot\hat{\bf P}_{\rm b}},
} 
where $\hat{{\bf x}}$ is the position operator of the impurity and $\hat{\bf P}_{\rm b}$ is the total momentum operator of bath phonons. After employing the transformation, the conserved quantity becomes the momentum operator $\hat{\bf p}$ of the impurity as inferred from the relation:
\eqn{\label{LLPconservation}
\hat{U}_{\rm LLP}^{\dagger}(\hat{\bf p}+\hat{{\bf P}}_{\rm b})\hat{U}_{\rm LLP}=\hat{\bf p}.
} 
In the transformed frame, $\hat{{\bf p}}$ can be taken as a classical variable and thus, the impurity is decoupled from the bath degrees of freedom, or said differently, its dynamics is completely frozen. This observation naturally motivates the following family of variational states:

\eqn{\label{LLPvar}
|\Psi_{\rm tot}(\xi)\rangle=\hat{U}_{\rm LLP}|{\bf p}\rangle|\Psi_{\rm b}(\xi)\rangle,
}
where $|{\bf p}\rangle$ is a single-particle state of the impurity having eigenmomentum ${\bf p}$, and $|\Psi_{\rm b}(\xi)\rangle$ is a bath wavefunction characterized by a set of variational parameters $\xi$. For example, when the bath is given by an ensemble of bosonic excitations, $|\Psi_{\rm b}\rangle$ can be chosen as a factorizable coherent state, as has been suggested in the original papers by Tomonaga  \cite{TS47} and Lee, Low and Pines \cite{LLP53}.  Variational approach combining the transformation~(\ref{LLP}) with a variety of efficiently parametrizable wavefunctions has been very successful in solving both in- and out-of-equilibrium problems and thus laid the cornerstone of successive studies in polaron physics  \cite{DJT09,DJT10,AAS_book,FRP55,BJ67,NP90,ZWM90,AT93,TJ09,NA10,CW11,CW112,RS13,VJ13,VJ14,LW14,FG15,SYE16,SYE162,YA17}.   

Our aim is to generalize this variational approach to yet another important paradigm in many-body physics: localized spin-impurity models (SIM) (Fig.~\ref{fig_schem1}). The most fundamental problem in SIM is the Kondo model \cite{KJ64}, a localized spin-1/2 impurity interacting with a fermionic bath. Its equilibrium properties are now theoretically well understood based on the results of the  perturbative renormalization group (RG) \cite{APW70}, Wilson's numerical renormalization group (NRG) \cite{WKG75,BR03,BL07,BRC08,SH08,BL09,BCA10} and the exact solution via the Bethe ansatz  \cite{PBW80,AND80,ANL81,NK81,AN83,SP89}. Yet, its out-of-equilibrium dynamics is still an area of active experimental   \cite{RL17,DFS02,THE11,LC11,IZ15,MMD17} and theoretical research  \cite{STL08,WP09,SM09,WP10,CG13,NP99,KA00,AR03,HA08,KM01,PM10,HA09L,HA09B,CT11,BS14,FS15,BCZ17,WSR04,SP04,AHKA06,DSLG08,WA09,HMF09,HMF10,NHTM17,AFB05,AFB06,AFB07,AFB08,RD08,JE10,LB14,NM15,DB17,LF96,LF98,SA98,LD05,VR13,SG14,MM13,BCJ16} with many open questions. Earlier works include real-time Monte Carlo   \cite{STL08,WP09,SM09,WP10,CG13}, perturbative RG  \cite{NP99,KA00,AR03,HA08,KM01,PM10} and Hamiltonian RG method  \cite{HA09L,HA09B,CT11}, coherent-state expansion \cite{BS14,FS15,BCZ17}, density-matrix renormalization group (DMRG) \cite{WSR04,SP04,AHKA06,DSLG08,WA09,HMF09,HMF10,NHTM17}, time-dependent NRG (TD-NRG) \cite{AFB05,AFB06,AFB07,AFB08,RD08,JE10,LB14}, time evolving decimation (TEBD) \cite{NM15,DB17}, and analytical solutions \cite{LF96,LF98,SA98,LD05,VR13,SG14,MM13,BCJ16}.

In spite of the rich theoretical toolbox for studying the quantum impurity systems, analysis of the long-time dynamics remains very challenging. Most theoretical approaches suffer from the same fundamental limitation: they become increasingly costly in terms of computational resources at long times. For example, it has been pointed out in TD-NRG that the logarithmic discretization may cause artifacts in predicting long-time dynamics  \cite{RA12}, and calculations based on the matrix-product states (MPS) are known to be extremely challenging in the long-time regimes because the large amount of entanglement forces one to use an exponentially large bond dimension \cite{SU11}. Another difficulty intrinsic to some of the methods is to understand spatiotemporal dynamics of the bath degrees of freedom since the latter are often integrated out or represented as a simplified effective bath, in which details of the microscopic eigenstates are omitted. Furthemore, the currently available methods have been constructed for a specific class of the Kondo impurity model, in which electrons move ballistically and interaction between the impurity spin and fermions is local. Extending these techniques to the case of electrons strongly scattered by disorder and nonlocal interactions with the impurity spin is not obvious. These challenges motivate us to develop a new theoretical approach to quantum impurity systems. In the accompanying paper \cite{YA18L}, we introduce a new canonical transformation that is the core of the proposed variational approach and provide strong evidence for the validity of our approach to correctly describe in- and out-of-equilibrium properties of SIM.

In this paper, we present comprehensive details of the proposed variational approach. For the sake of completeness, we first provide the construction of the new transformation that generates the entanglement between the impurity and the bath. In contrast to the LLP transformation~(\ref{LLP}) of going into the comoving frame of the mobile spinless impurity, such a construction of the canonical transformation in SIM is not obvious due to the SU(2)-commutation relation of the impurity spin operators. We discuss a general approach to constructing such transformations in this paper.
We then combine the transformation with Gaussian states to obtain a family of variational states that can efficiently capture the impurity-bath entanglement. We provide a set of nonlinear equations of motions for the covariance matrix to study ground-state and out-of-equilibrium properties of generic SIM. We benchmark our approach by applying it to the anisotropic Kondo model and compare results with those obtained using MPS ansatz. We also compare our results to the ones obtained from the Yosida ansatz and the exact solution via the Bethe ansatz. We analyze out-of-equilibrium dynamics and transport properties in the two-lead Kondo model and demonstrate that our approach can be used to compute the long-time spatiotemporal dynamics and the conductance at finite bias and magnetic field.  The obtained results are consistent with the previous studies in the Anderson model  \cite{KRM02,AHKA06,JE10} and the exact solutions at the Toulouse point  \cite{BCJ16}.

This paper is organized as follows. In Sec.~\ref{secformalism}, we present a general concept of our variational approach to  SIM. In particular, we discuss the canonical transformation introduced in Ref.~\cite{YA18L} that decouples the impurity from the bath degrees of freedom. We then derive the equations of motions for the covariance matrix of fermionic Gaussian states that describe the ground-state properties and real-time evolutions of SIM. In Sec.~\ref{secKondo}, we apply our theory to the anisotropic Kondo model and benchmark it with MPS-based calculations and the exact solution via the Bethe ansatz. In Sec.~\ref{secKondotwo}, we analyze transport dynamics in the two-lead Kondo model in further detail and reveal long-time spatiotemporal dynamics and the conductance behavior at finite bias and magnetic fields. Finally, we summarize the results in Sec.~\ref{secConclusion}.

\section{General Formalism}\label{secformalism}
\subsection{Canonical transformation}
We first formulate our variational approach to SIM in a general way. The key idea is to introduce a new canonical transformation that completely decouples the impurity and the bath degrees of freedom for a generic spin-1/2 impurity Hamiltonian $\hat{H}$ by employing its parity symmetry $\hat{\mathbb P}$, i.e., $[\hat{H},\hat{\mathbb P}]=0$ with $\hat{\mathbb P}^{2}=1$.
Specifically, we aim to find a unitary transformation $\hat{U}$ that maps this parity operator $\hat{\mathbb P}$ to an impurity operator as (c.f. Eq.~(\ref{LLPconservation}))
\eqn{\label{Conserve}
\hat{U}^{\dagger}\hat{\mathbb P}\hat{U}=\mathbf{n}\cdot\boldsymbol{\hat{\sigma}}_{\rm imp},
}
where $\hat{\boldsymbol \sigma}_{\rm imp}=(\hat{\sigma}_{\rm imp}^{x},\hat{\sigma}_{\rm imp}^{y},\hat{\sigma}_{\rm imp}^{z})^{\rm T}$ is a vector of the impurity spin-1/2 operator  and  $\bf n$ is some specific direction defined by a three dimensional real vector. It then follows that in the transformed frame the Hamiltonian commutes with the impurity operator $[\hat{U}^{\dagger}\hat{H}\hat{U},\mathbf{n}\cdot\boldsymbol{\hat{\sigma}}_{\rm imp}]=0$ such that the impurity dynamics is frozen, i.e., the transformed Hamiltonian conditioned on a classical variable $\mathbf{n}\cdot\boldsymbol{\hat{\sigma}}_{\rm imp}=\pm1$ only contains the bath degrees of freedom. The price that one pays for decoupling the impurity spin is the appearance of nonlocal multiparticle interactions between the bath degrees of freedom. In spite of the seeming complexity of the fermionic interactions in the decoupled frame,  we show that their dynamics can be efficiently described using Gaussian variational wavefunctions, which require the number of parameters that grows at most polynomially with the system size.

In this paper, we apply this approach to a spin-1/2 impurity interacting with a fermionic bath:

\eqn{\label{Ham}
\hat{H}&=&\sum_{lm\alpha}h_{lm}\hat{\Psi}_{l\alpha}^{\dagger}\hat{\Psi}_{m\alpha}-h_z\hat{s}_{\rm imp}^{z}+\hat{\bf s}_{\rm imp}\cdot\hat{\bf \Sigma},
}
where $\hat{\Psi}_{l\alpha}^{\dagger}$ ($\hat{\Psi}_{l\alpha}$) is a fermionic creation (annihiliation) operator corresponding to a bath mode $l=1,2,\ldots,N_{f}$ and spin-$z$ component $\alpha=\uparrow,\downarrow$, $h_{lm}$ is an arbitrary $N_f\times N_f$ Hermitian matrix describing a single-particle Hamiltonian of a bath, and $h_z$ is a magnetic field acting on the impurity. We define the impurity spin operator $\hat{\bf s}_{\rm imp}=\hat{\boldsymbol \sigma}_{\rm imp}/2$ and introduce the bath-spin density operator including couplings as
\eqn{\label{impbath}
\hat{\Sigma}^{\gamma}=\frac{1}{2}\sum_{lm\alpha\beta}g_{lm}^{\gamma}\hat{\Psi}_{l\alpha}^{\dagger}\sigma_{\alpha\beta}^{\gamma}\hat{\Psi}_{m\beta},
}
where $\gamma=x,y,z$.
The first term in Eq.~(\ref{Ham}) describes a noninteracting bath, the second term describes a local magnetic effect acting on the impurity, and the third term characterizes the interaction between the impurity and the bath. The impurity-bath  couplings in Eq.~(\ref{impbath}) are determined by $N_f\times N_f$ Hermitian matrices  $g_{lm}^{\gamma}$ labeled by $\gamma=x,y,z$ and can in general be anisotropic and long-range.  While the interaction leads to strong impurity-bath entanglement, we note that it can also generate entanglement between bath modes since the ones that interact with the impurity are not diagonal with respect to the eigenbasis of bath Hamiltonian $h_{lm}$ in general. The magnetic-field term acting on the bath can be also included; we omit that for simplicity. 

Our canonical transformation relies on the parity symmetry hidden in the Hamiltonian~(\ref{Ham}). To unveil this symmetry, we introduce the operator $\hat{\mathbb{P}}=\hat{\sigma}_{\rm imp}^{z}\hat{\mathbb P}_{\rm bath}$ with a bath parity operator
\eqn{\label{pbath}
\hat{\mathbb P}_{\rm bath}=e^{(i\pi/2)(\sum_{l}\hat{\sigma}^{z}_{l}+\hat{N})}=e^{i\pi\hat{N}_{\uparrow}},
} 
where $\hat{N}$ is the total particle number in the bath, $\hat{\sigma}_l^{\gamma}=\sum_{\alpha\beta}\hat{\Psi}^{\dagger}_{l\alpha}\sigma^{\gamma}_{\alpha\beta}\hat{\Psi}_{l\beta}$ ($\gamma=x,y,z$) is a spin-density operator with a bath mode $l$, and $\hat{N}_{\uparrow}$ is the number of spin-up fermions. These operators satisfy $\hat{\mathbb P}_{\rm bath}^{2}=\hat{\mathbb P}^{2}=1$. We observe that the Hamiltonian~(\ref{Ham}) conserves $\hat{\mathbb P}$, i.e., $[\hat{H},\hat{\mathbb P}]=0$. This corresponds to the symmetry under the rotation of the entire system around $z$ axis by $\pi$, which maps both impurity and bath spins as $\hat{\mathbb P}^{-1}\hat{\sigma}^{x,y}\hat{\mathbb P}= -\hat{\sigma}^{x,y}$ while it keeps  $\hat{\mathbb P}^{-1}\hat{\sigma}^{z}\hat{\mathbb P}= \hat{\sigma}^{z}$. 

We employ this parity conservation to construct the disentangling transformation $\hat{U}$ satisfying 
\eqn{
\hat{U}^{\dagger}\hat{\mathbb P}\hat{U}=\hat{\sigma}_{\rm imp}^{x}
} 
such that the impurity spin turns out to be a conserved quantity in the transformed frame. Here we choose ${\bf n}=(1,0,0)^{\rm T}$ in Eq.~(\ref{Conserve}); other choices will lead to the same class of variational states. We define a unitary transformation $\hat{U}$ as 
\eqn{
\hat{U}=\exp\left[\frac{i\pi}{4}\hat{\sigma}_{\rm imp}^{y}\hat{\mathbb P}_{\rm bath}\right]=\frac{1}{\sqrt{2}}\left(1+i\hat{\sigma}_{\rm imp}^{y}\hat{\mathbb P}_{\rm bath}\right).
}
Employing this transformation, we arrive at the following transformed Hamiltonian $\hat{\tilde{H}}\!  =\!\hat{U}^{\dagger}\hat{H}\hat{U}$:
 
\eqn{
\hat{\tilde{H}} \! &=&\!\sum_{lm\alpha}h_{lm}\hat{\Psi}_{l\alpha}^{\dagger}\hat{\Psi}_{m\alpha}-h_z\hat{s}_{\rm imp}^{x}\hat{{\mathbb P}}_{\rm bath}\nonumber\\
&&+\hat{s}_{\rm imp}^{x}\hat{\Sigma}^{x}
+\hat{{\mathbb P}}_{\rm bath}\left(-\frac{i\hat{\Sigma}^{y}}{2}+\hat{s}_{\rm imp}^{x}\hat{\Sigma}^{z}\right).
\label{transformedHam}
}
As expected from the construction, the impurity spin commutes with the Hamiltonian $[\hat{\tilde{H}},\hat{s}_{\rm imp}^x]=0$, and thus we can take the impurity operator as a conserved number $\sigma_{\rm imp}^{x}=2s_{\rm imp}^{x}=\pm1$  in the transformed frame. Decoupling of the impurity spin came at the cost of introducing interactions among the bath particles. Note that these interactions are multiparticle and nonlocal, as can be seen from the form of the operator $\hat{\mathbb P}_{\rm bath}$. The appearance of the bath interaction can be interpreted as spin exchange between fermions via the impurity spin. This is analogous to the case of the mobile spinless impurity \cite{LLP53}, where a nonlocal phonon-phonon interaction is introduced after transforming to the comoving frame of the impurity via the LLP transformation.  We emphasize that the elimination of the impurity  relies only on the elemental parity symmetry in the original Hamiltonian and thus should have a wide applicability. The construction of $\hat{U}$ holds true for arbitrary conserved parity, and can be also applied to two-impurity systems \cite{SP18}.

It follows from Eq.~(\ref{transformedHam}) that for the even bath-particle number $N$ and the zero magnetic field $h_z=0$, the Hamiltonian has two degenerate equivalent energy sectors corresponding to the conserved quantity $\sigma_{\rm imp}^{x}=\pm1$ in the transformed frame. This is because the two sectors in Eq.~(\ref{transformedHam}) can be exactly related via the additional unitary transformation $\hat{U}^{y}_{\rm bath}=e^{(i\pi/2)\sum_{l}\hat{\sigma}_{l}^{y}}$, which maps the bath spins as $\hat{\sigma}^{x,z}_{l}\to -\hat{\sigma}^{x,z}_{l}$. To our knowledge, such an exact spectrum degeneracy hidden in generic SIM has not been pointed out except for some specific cases that are exactly solvable via the Bethe ansatz \cite{PBW80,AND80,ANL81,NK81,AN83,SP89} or at the Toulouse point \cite{ZG00}. 

The explicit form of variational states is (c.f. Eq.~(\ref{LLPvar}))
\eqn{\label{varoriginal}
|\Psi\rangle&=&\hat{U}|\pm_{x}\rangle_{\rm imp}|\Psi_{\rm b}\rangle\nonumber\\
&=&|\!\uparrow~\!\!\rangle_{\rm imp}\hat{\mathbb P}_{\pm}|\Psi_{\rm b}\rangle\pm|\!\downarrow~\!\!\rangle_{\rm imp}\hat{\mathbb P}_{\mp}|\Psi_{\rm b}\rangle,
} 
where $|\pm_{x}\rangle_{\rm imp}$ represents the eigenstate of  $\hat{\sigma}_{\rm imp}^{x}$ with eigenvalue $\pm 1$, $|\Psi_{\rm b}\rangle$ represents a bath wavefunction, and $\hat{{\mathbb P}}_{\pm}=(1\pm\hat{{\mathbb P}}_{\rm bath})/2$ is the projection onto the subspace with even or odd number of spin-up fermions. As we will see later, this form of variational wavefunction can naturally capture the strong impurity-bath entanglement, which is an essential feature of, for example, the formation of the Kondo singlet. 

We stress that our variational approach is unbiased in the sense that the variational states~(\ref{varoriginal}) are constructed without any particular {\it a priori} knowledge about the underlying impurity physics such as the Kondo physics. The canonical transformation $\hat{U}$ includes neither parameters in the Hamiltonian (e.g., Kondo coupling) nor any variational parameters. While we will choose $|\Psi_{\rm b}\rangle$ as the Gaussian states in the next subsection, the variational states~(\ref{varoriginal}) are still unbiased because the Gaussian states take into account all the two-particle excitations in an unbiased manner.

\subsection{Variational time-evolution equations}
\subsubsection{Fermionic Gaussian states}
To solve SIM efficiently, we have to introduce a family of variational bath wavefunctions that can approximate the ground state and real-time evolutions governed by the Hamiltonian~(\ref{transformedHam}) while they are simple enough so that calculations can be done in a tractable manner. In the decoupled frame, we choose variational many-body states for the bath as fermionic Gaussian states \cite{WC12,MJ13,CVK10}.
It is convenient to introduce the Majorana operators $\hat{\psi}_{1,l\alpha}=\hat{\Psi}^{\dagger}_{l\alpha}+\hat{\Psi}_{l\alpha}$ and $\hat{\psi}_{2,l\alpha}=i(\hat{\Psi}^{\dagger}_{l\alpha}-\hat{\Psi}_{l\alpha})$ satisfying the anticommutation relation $\{ \hat{\psi}_{\xi,l\alpha},\hat{\psi}_{\eta,m\beta}\} =2\delta_{\xi\eta}\delta_{lm}\delta_{\alpha\beta}$ with $\xi,\eta=1,2$.
We describe the bath by a pure fermionic Gaussian state $|\Psi_{\rm G}\rangle$ that is fully characterized by its $4N_f\!\times \!4N_f$ covariance matrix $\Gamma$  \cite{WC12,MJ13,CVK10}: 
\eqn{
\Gamma=\frac{i}{2}\left\langle[\hat{\boldsymbol \psi},\hat{\boldsymbol \psi}^{\rm T}]\right\rangle_{\rm G}, 
}
where $(\Gamma)_{ij}=-(\Gamma)_{ji}\in\mathbb{R}$ is real antisymmetric  and $\Gamma^{2}=-{\rm I}_{4N_{f}}$ for pure states with ${\rm I}_{d}$ being the $d\times d$ unit matrix. Here, $\langle\cdots\rangle_{\rm G}$ denotes an expectation value with respect to the Gaussian state $|\Psi_{\rm G}\rangle$, and we introduced the Majorana operators $\hat{\boldsymbol{\psi}}=(\hat{\boldsymbol{\psi}}_{1},\hat{\boldsymbol{\psi}}_{2})^{\rm T},$
where we choose the ordering of a row vector $\hat{\boldsymbol \psi}_{\xi}$ with $\xi=1,2$ as
\eqn{
\hat{\boldsymbol{\psi}}_{\xi}&=&(\hat{\psi}_{\xi,1\uparrow},\ldots,\hat{\psi}_{\xi,N_f\uparrow},\hat{\psi}_{\xi,1\downarrow},\ldots,\hat{\psi}_{\xi,N_f\downarrow}).
}
We can explicitly write the bath state as
\begin{equation}
|\Psi_{\rm G}\rangle=e^{\frac{1}{4}\hat{\boldsymbol{\psi}}^{\rm T}X\hat{\boldsymbol{\psi}}}|0\rangle\equiv\hat{U}_{{\rm G}}|0\rangle,
\end{equation}
where we define the Gaussian unitary operator by $\hat{U}_{\rm G}$ and  introduce a real-antisymmetric matrix $(X)_{ij}=-(X)_{ji}\in\mathbb{R}$. The latter can be related to $\Gamma$ via
\begin{equation}
\label{GmXi}
\Gamma=-\Xi\,\sigma\left(\Xi\right)^{\rm T},
\end{equation}
where $\Xi=e^{X}$ and $\sigma=i\sigma^{y}\otimes{\rm I}_{2N_f}$.
It will be also useful to define the $2N_f\!\times \!2N_f$ correlation matrix:
\eqn{
\Gamma_{f}=\langle\hat{\boldsymbol \Psi}^{\dagger}\hat{\boldsymbol \Psi}\rangle_{{\rm G}}
}
in terms of Dirac fermions
\eqn{
\hat{\boldsymbol \Psi}=(\hat{\Psi}_{1\uparrow},\ldots,\hat{\Psi}_{N_f\uparrow},\hat{\Psi}_{1\downarrow},\ldots,\hat{\Psi}_{N_f\downarrow}).
}
\subsubsection{Imaginary- and real-time evolutions}

We approximate the exact time evolution of a bath wavefunction $|\Psi_{\rm b}\rangle$ by projecting it onto the manifold spanned by the family of variational states. This can be done by employing the time-dependent variational principle  \cite{JR79,KP08,CVK10,ST17}, which allows us to study ground-state properties via the imaginary-time evolution and also out-of-equilibrium dynamics via the real-time evolution. Let us first formulate the former one.

The imaginary-time evolution 
\eqn{\label{imagint}
|\Psi_{\rm b}(\tau)\rangle=\frac{e^{-\hat{\tilde{H}}\tau}|\Psi_{\rm b}(0)\rangle}{\left\Vert e^{-\hat{\tilde{H}}\tau}|\Psi_{\rm b}(0)\rangle\right\Vert}
}
gives the ground state in the asymptotic limit $\tau\to\infty$ if the initial seed state $|\Psi_{\rm b}(0)\rangle$ has a non-zero overlap with the ground state.
Differentiating Eq.~(\ref{imagint}), we obtain the equation of motion
\begin{equation}
\frac{d}{d\tau}|\Psi_{\rm b}(\tau)\rangle=-(\hat{\tilde{H}}-E)|\Psi_{\rm b}(\tau)\rangle,
\end{equation}
where $E=\langle\Psi_{\rm b}(\tau)|\hat{\tilde{H}}|\Psi_{\rm b}(\tau)\rangle$ represents the mean energy. If we consider the covariance matrix $\Gamma$ of the Gaussian state $|\Psi_{\rm G}\rangle$ to be the time-dependent variational parameters, their imaginary-time evolution equation can be obtained by minimizing its deviation $\epsilon$ from the exact imaginary-time evolution:

\eqn{\label{error}
\epsilon=\left\Vert \frac{d}{d\tau}|\Psi_{\rm G}(\tau)\rangle+(\hat{\tilde{H}}-E_{\rm var})|\Psi_{\rm G}(\tau)\rangle\right\Vert ^{2},\nonumber\\
}
where $E_{\rm var}=\langle\Psi_{\rm G}(\tau)|\hat{\tilde{H}}|\Psi_{\rm G}(\tau)\rangle$ is the variational energy.
This is formally equivalent to solving the projected differential equation:
\begin{equation}
\label{vareqformal}
\frac{d}{d\tau}|\Psi_{\rm G}(\tau)\rangle=-\hat{{\rm P}}_{\partial\Gamma}(\hat{\tilde{H}}-E_{\rm var})|\Psi_{\rm G}(\tau)\rangle,
\end{equation}
where $\hat{\rm P}_{\partial\Gamma}$ is the projector onto the subspace  spanned by tangent vectors of the variational manifold. On  one hand, the left-hand side of Eq.~(\ref{vareqformal}) gives

\eqn{
\frac{d}{d\tau}|\Psi_{\rm G}(\tau)\rangle
\!=\!\hat{U}_{{\rm G}}\left(\!\frac{1}{4}\!:\hat{\boldsymbol{\psi}}^{\rm T}\Xi^{\rm T}\frac{d\Xi}{d\tau}\hat{\boldsymbol{\psi}}:\!+\frac{i}{4}{\rm Tr}\left[\Xi^{\rm T}\frac{d\Xi}{d\tau}\Gamma\right]\right)|0\rangle,\nonumber\\\label{leftvar}
}
where $:\;:$ represents taking the normal order of the Dirac operators $\hat{\Psi}$ and $\hat{\Psi}^{\dagger}$. On the other hand, we can write the right-hand side of Eq.~(\ref{vareqformal}) as:
\eqn{
-(\hat{\tilde{H}}-E_{\rm var})|\Psi_{\rm G}(\tau)\rangle=-\left(\frac{i}{4}:\hat{\boldsymbol{\psi}}^{\rm T}\Xi^{\rm T}{\cal H}\Xi\hat{\boldsymbol{\psi}}:+\delta\hat{O}\right)|0\rangle,\nonumber\\
\label{rightvar}
}
where ${\cal H}=4\delta E_{\rm var}/\delta\Gamma$ is the functional derivative  of the variational energy   \cite{ST17}  and $\delta \hat{O}$ denotes the cubic and higher order contributions of $\hat{\boldsymbol \psi}$ that are orthogonal to the tangential space and thus will be projected out by $\hat{\rm P}_{\partial \Gamma}$ in Eq.~(\ref{vareqformal}). Comparing Eqs.~(\ref{leftvar}) and (\ref{rightvar}), and using Eq.~(\ref{GmXi}), we can uniquely determine the imaginary time-evolution equation of the covariance matrix $\Gamma$ as \cite{CVK10,ST17}
\eqn{\label{imagvar}
\frac{d\Gamma}{d\tau}&=&-{\cal H}-\Gamma {\cal H}\Gamma,
}
which guarantees that the variational energy $E_{\rm var}$  monotonically decreases and the variational ground state is achieved in the limit $\tau\to\infty$. In this limit, the error $\epsilon$ in Eq.~(\ref{error}) is equivalent to the variance of the energy in the reached ground state and can be used as an indicator to check the accuracy of the variational state. 

In the similar way, we can derive the equation of motion for $\Gamma$ in the real-time evolution
\eqn{
|\Psi_{\rm b}(t)\rangle=e^{-i\hat{\tilde{H}}t}|\Psi_{\rm b}(0)\rangle.
}
The projection
\begin{equation}
\frac{d}{dt}|\Psi_{\rm G}(t)\rangle=-i\hat{{\rm P}}_{\partial\Gamma}\hat{\tilde{H}}|\Psi_{\rm G}(t)\rangle
\end{equation}
of the Schr{\"o}dinger equation on the variational manifold leads to the following real-time evolution equation of the covariance matrix $\Gamma$ \cite{CVK10,ST17}:
\eqn{\label{realvar}
\frac{d\Gamma}{dt}&=&{\cal H}\Gamma-\Gamma {\cal H}.
}

\subsection{Functional derivative of variational energy}
To analytically establish the variational time-evolution equations (\ref{imagvar}) and (\ref{realvar}) of the covariance matrix $\Gamma$, we have to obtain the functional derivative ${\cal H}=4\delta E_{\rm var}/\delta\Gamma$ of the variational energy. In this subsection, we give its explicit analytical expression. First of all, we write the expectation value $E_{\rm var} =\langle\hat{\tilde{H}}\rangle_{{\rm G}}$ of the Hamiltonian (\ref{transformedHam}) with respect to the fermionic Gaussian state as

\eqn{
E_{\rm var}  \!& =&\!\sum_{lm\alpha}h_{lm}(\Gamma_{f})_{l\alpha,m\alpha}
 -\!\frac{h_z}{2}\sigma_{\rm imp}^{x}\langle\hat{\mathbb{P}}_{{\rm bath}}\rangle_{{\rm G}}\nonumber\\
& +&\frac{1}{4}\sum_{lm\alpha\beta}g_{lm}^{x}\sigma_{{\rm imp}}^{x}\sigma_{\alpha\beta}^{x}(\Gamma_{f})_{l\alpha,m\beta}\nonumber\\
\!&+&\!\frac{1}{4}\!\sum_{lm\alpha\beta}\!\!\left(-ig_{lm}^{y}\sigma^{y}\!+g_{lm}^{z}\sigma_{{\rm imp}}^{x}\sigma^{z}\right)_{\alpha\beta}\!\!(\Gamma_{f}^{\rm P})_{l\alpha,m\beta},\! 
\label{meanE}
}
where we condition the impurity operator on a classical number $\sigma_{\rm imp}^{x}=\pm1$ and introduce  the $2N_f\!\times\! 2N_f$ matrix containing the parity operator as
$\Gamma_{f}^{\rm P}=\langle\hat{\mathbb{P}}_{{\rm bath}}\hat{\boldsymbol\Psi}^{\dagger}\hat{\boldsymbol\Psi}\rangle_{{\rm G}}.$
The values of $\langle\hat{\mathbb P}_{\rm bath}\rangle_{\rm G}$ and $\Gamma_{f}^{\rm P}$ can be obtained as
\eqn{\label{parityG}
\langle\hat{\mathbb{P}}_{{\rm bath}}\rangle_{\rm G}&=&(-1)^{N_{f}}{\rm Pf}\left[\frac{\Gamma_{F}}{2}\right],
}
and
\eqn{
\Gamma _{f}^{\mathrm{P}}&&=\frac{1}{4}\langle \hat{
\mathbb{P}}_{\mathrm{bath}}\rangle _{\mathrm{G}}\nonumber\\
&&\times\Sigma _{z}(\mathrm{I}
_{2N_{f}},-i\mathrm{I}_{2N_{f}})\Upsilon ^{-1}\! \left( \Gamma \sigma -\mathrm{I}_{4N_{f}}\right) \!\!\left(\!
\begin{array}{c}
\mathrm{I}_{2N_{f}} \\
i\mathrm{I}_{2N_{f}}
\end{array}
\!\right),
}
where $\rm Pf$ denotes the Pfaffian and the matrices 
\eqn{
\Gamma_{F}&=&\sqrt{{\rm I}_{4N_f}+\Lambda}\,\Gamma\sqrt{{\rm I}_{4N_f}+\Lambda}-({\rm I}_{4N_f}-\Lambda)\sigma,\\
\Upsilon  &=&\mathrm{I}_{4N_{f}}+\frac{1}{2}\left( \Gamma \sigma -\mathrm{I}%
_{4N_{f}}\right) \left( \mathrm{I}_{4N_{f}}+\Lambda \right),
}
are defined by $\Lambda={\rm I}_2\otimes\Sigma_z$ and 
$\Sigma_z=\sigma^z\otimes{\rm I}_{N_f}$. We recall that the matrix $\sigma$ is defined below Eq.~(\ref{GmXi}).

Since the first and third terms in the Hamiltonian~(\ref{transformedHam}) are quadratic, we can write them exactly in the Majorana basis as 
$i\hat{\boldsymbol{\psi }}^{\rm T}\mathcal{H}_{0}\hat{\boldsymbol{\psi }}/4$ with
\eqn{
\mathcal{H}_{0}=i\sigma ^{y}\otimes \lbrack \mathrm{I}_{2}\otimes h_{lm}+(\sigma _{\mathrm{imp}}^{x}/4)\,\sigma ^{x}\otimes g_{lm}^{x}],
}
where $h_{lm}$ and $g_{lm}^{x}$ are understood to be $N_f\!\times\! N_f$ real matrices. Thus, the functional derivative ${\cal H}=4\delta E_{\rm var}/\delta\Gamma$ of the mean energy~(\ref{meanE}) is given by
\eqn{\label{interH}
\mathcal{H}\!=\!\mathcal{H}_{0}+\frac{\delta}{\delta\Gamma}\left[-2h_{z}\sigma_{\mathrm{imp}}^{x}\langle \hat{{\mathbb{P}}}_{\mathrm{bath}}\rangle
_{\mathrm{G}}+{\rm Tr}({\cal S}^{\rm T}\Gamma_{f}^{\rm P})\right],
}
where we introduce the matrix
\eqn{
\mathcal{S}=-i\sigma ^{y}\otimes g_{lm}^{y}+\sigma_{\mathrm{imp}%
}^{x}\sigma ^{z}\otimes g_{lm}^{z}.
}
Taking the derivatives of $\langle\hat{\mathbb P}_{\rm bath}\rangle_{\rm G}$ and $\Gamma_{f}^{\rm P}$ with respect to the covariance matrix $\Gamma$ in Eq.~(\ref{interH}), we finally obtain the analytical expression of the functional derivative $\cal H$:

\eqn{\label{hm}
\mathcal{H} \!&=&\!\mathcal{H}_{0}\!+\!\left[h_z\sigma _{\mathrm{imp}}^{x}\langle
\hat{{\mathbb{P}}}_{\mathrm{bath}}\rangle_{\mathrm{G}}\!-\!
\frac{1}{2}{\rm Tr}(\mathcal{S}^{\rm T}\Gamma _{f}^{\mathrm{P}})\right]\!\mathcal{P}\nonumber\\
&&-\frac{i}{4}\langle \hat{{\mathbb{P}}}_{\mathrm{bath}}\rangle
_{\mathrm{G}}\;\mathcal{A}\left[ {\cal V}\mathcal{S}^{\rm T}\Sigma_{z}{\cal V}^{\dagger}\right], 
}
where ${\cal A}[M]\!=\!(M\!-\!M^{\rm T})/2$ denotes the matrix antisymmetrization and we introduce the matrices
\eqn{
\mathcal{P}&=&\sqrt{\mathrm{I}_{4N_{f}}+\Lambda }\Gamma _{F}^{-1}\sqrt{\mathrm{I}_{4N_{f}}+\Lambda},\\
{\cal V}&=&(\Upsilon^{\rm T})^{-1}\left(\begin{array}{c}
{\rm I}_{2N_f}\\
i{\rm I}_{2N_f}
\end{array}\right).
}
Integrating Eqs.~(\ref{imagvar}) and (\ref{realvar}) with the general form~(\ref{hm}) of the functional derivative, one can study ground-state properties and out-of-equilibrium dynamics of  SIM on demand.

 \section{Application to the anisotropic Kondo model}\label{secKondo}
 \subsection{Model}
We benchmark our general variational approach by applying it to the anisotropic Kondo model and comparing the results with those obtained using matrix-product state (MPS)  \cite{FV08}. We also compare our results to the Yosida ansatz \cite{KY66} and the exact solution via the Bethe ansatz \cite{PBW80,AND80,ANL81,NK81,AN83,SP89}. The one-dimensional Kondo Hamiltonian is given by
 \eqn{\label{anisotropicK}
\hat{H}_{\rm K}\!=\!-t_{\rm h}\!\!\sum_{l=-L}^{L}\left(\hat{c}^{\dagger}_{l\alpha}\hat{c}_{l+1\alpha}\!+\!{\rm h.c.}\right)\!+\!\frac{1}{4}\!\sum_{\gamma}\!J_{\gamma}\hat{\sigma}^{\gamma}_{\rm imp}\hat{c}_{0\alpha}^{\dagger}\sigma^{\gamma}_{\alpha\beta}\hat{c}_{0\beta},\nonumber\\
}
where $\hat{c}^{\dagger}_{l\alpha}$ ($\hat{c}_{l\alpha}$) creates (annhilates) a fermion with position $l$ and spin $\alpha$. The spin-1/2 impurity $\hat{\sigma}_{\rm imp}^{\gamma}$ locally interacts with particles at the impurity site $l=0$ via the anisotropic couplings $J_{x,y}=J_\perp$ and $J_z=J_{\parallel}$. We choose the unit $t_{\rm h}=1$ and the summations over $\alpha,\beta$ are understood to be contracted hereafter. This model shows a quantum phase transition  \cite{LAJ87} between an antiferromagnetic (AFM) phase and a ferromagnetic (FM) phase. The former leads to the formation of the singlet state between the impurity and bath spins, leading to the vanishing impurity magnetization $\langle\hat{\sigma}_{\rm imp}^{z}\rangle=0$. The latter exhibits the triplet formation and the impurity magnetization takes a non-zero, finite value in general.

We remark that, employing the infinite-bandwidth approximation and the bosonization, the Kondo model can be mapped to the spin-boson model \cite{LAJ87}. Within this treatment, the impurity magnetization and the spatiotemporal dynamics have been previously studied by TD-NRG method \cite{AFB06,LB14} and also by the bosonic Gaussian states combined with a unitary transformation \cite{ST17}. In the latter, the unitary transformation  was specifically designed to the spin-boson model and one had to choose a specific symmetry sector to obtain meaningful results. In contrast, we here construct a completely different, general family of variational states by introducing the new decoupling transformation $\hat{U}$. The transformation $\hat{U}$ specifies no specific conditions on a physical system as far as it contains (arbitrary) parity symmetry $\hat{\mathbb{P}}$ and a spin-1/2 operator. In view of this strong versatility, our approach can be applied to a much wider class of problems than the previous work \cite{ST17}. Moreover, we apply this general approach to analyze in and out of equilibrium problems of the fermonic Kondo models on the lattice, which are more challenging problems than the bosonized version due to their intrinsically finite bandwidth. 

To apply our general formalism in Sec.~\ref{secformalism}, we note that in the Hamiltonian~(\ref{anisotropicK}) only the symmetric bath modes couple to the impurity spin. We thus identify the fermionic bath modes $\hat{\Psi}$ as
\eqn{\label{compbasis}
\hat{\Psi}_{0\alpha}=\hat{c}_{0\alpha},\;\;\;\hat{\Psi}_{l\alpha}=\frac{1}{\sqrt{2}}\left(\hat{c}_{l\alpha}+\hat{c}_{-l\alpha}\right)
}
with $l=1,2,\ldots,L$. The corresponding bath Hamiltonian $h_{lm}$ is given by the following $(L\!+\!1)\!\times\!(L\!+\!1)$ hopping matrix $h_1$ of the single lead:
\eqn{\label{hopping1}
h_1=(-t_{\rm h})\left(
\begin{array}{ccccc}
0 & \sqrt{2} & 0 & \cdots  & 0 \\
\sqrt{2} & 0 & 1 & 0 & \vdots  \\
0 & 1 & 0 & 1 & \vdots  \\
\vdots  & 0 & 1 & \ddots  & 1 \\
0 & \cdots  & \cdots  & 1 & 0%
\end{array}%
\right).
}
The couplings $g_{lm}^{\gamma}$ are identified as the local Kondo interaction $J_{\gamma}\delta_{l0}\delta_{m0}$. In this section, we set the magnetic field to be zero $h_z=0$ for simplicity. 

Solving Eqs.~(\ref{imagvar}) and (\ref{realvar}) with the functional derivative~(\ref{hm}), we obtain the covariance matrix $\Gamma$ corresponding to the ground state and the real-time dynamics. 
The associated observables can be efficiently calculated in terms of the covariance matrix. For example, the impurity-bath spin correlations for each direction are obtained from 

\begin{equation}
\chi^{x}_{l}=\frac{1}{4}\langle \hat{\sigma}_{{\rm imp}}^{x}\hat{\sigma}_{l}^{x}\rangle =\frac{1}{4}\sigma_{{\rm imp}}^{x}\sigma^x_{\alpha\beta}(\Gamma_{f})_{l\alpha,l\beta},
\end{equation}
\begin{equation}
\chi^{y}_{l}=\frac{1}{4}\langle \hat{\sigma}_{{\rm imp}}^{y}\hat{\sigma}_{l}^{y}\rangle =\frac{1}{4}(-i\sigma^{y}_{\alpha\beta})(\Gamma_{f}^{\rm P})_{l\alpha,l\beta},
\end{equation}
and
\begin{equation}
\chi^{z}_{l}=\frac{1}{4}\langle \hat{\sigma}_{{\rm imp}}^{z}\hat{\sigma}_{l}^{z}\rangle =\frac{1}{4}\sigma_{{\rm imp}}^{x}\sigma^z_{\alpha\beta}(\Gamma_{f}^{\rm P})_{l\alpha,l\beta},
\end{equation}
where $\langle\cdots\rangle$ denotes an expectation value with respect to wavefuntion in the {\it original} frame. The impurity magnetization can be obtained from
\begin{equation}
\langle\hat{\sigma}^{z}_{\rm imp}\rangle=\sigma_{\rm imp}^{x}\langle \hat{\mathbb P}_{\rm bath}\rangle_{{\rm G}}=\sigma_{\rm imp}^{x}(-1)^{N_{f}}{\rm Pf}\left[\frac{\Gamma_{F}}{2}\right],
\end{equation}
where we use Eq.~(\ref{parityG}) in the second equality.

 \subsection{Structure of variational ground state}
In this subsection, we discuss how the entanglement in the Kondo-singlet state is naturally encoded in our variational ground state. It is useful to choose a correct sector of variational manifold by specifying an appropriate conserved quantum number. In particular, if the initial seed state is an eigenstate of the total spin-$z$ component $\hat{\sigma}^{z}_{\rm tot}=\hat{\sigma}_{\mathrm{imp}}^{z}+\hat{\sigma}_{\rm bath}^{z}$ with $\hat{\sigma}_{\rm bath}^{z}=\sum_{l=0}^{L}\hat{\sigma}_{l}^z$, its value $\sigma^{z}_{\rm tot}$ is conserved through the imaginary-time evolution due to $[\hat{\sigma}^{z}_{\rm tot},H]=0$.

To show the singlet behavior explicitly, we investigate the variational ground state by choosing the initial seed state in the sector $\sigma^{z}_{\rm tot}=0$. As inferred from the definition of the bath parity~(\ref{pbath}) and the variational ansatz (\ref{varoriginal}) conditioned on the sector $\sigma_{\rm imp}^{x}=1$, in the {\it original} frame the spin-up impurity $|\!\!\uparrow\rangle$ is coupled to the bath with even number of spin-up fermions $N_{\uparrow}$ (correspondingly, odd number of spin-down fermions $N_{\downarrow}=N_{\uparrow }+1$), while the spin-down impurity $|\!\!\downarrow\rangle$ is coupled to the bath having odd (even) number of spin-up (-down) fermions $N_{\uparrow}$ ($N_{\downarrow}=N_{\uparrow }-1$). It then follows that the projected states $|\Psi_{\rm \pm}\rangle=\hat{\mathbb P}_{\pm}|\Psi_{\rm b}\rangle$, which couple with the impurity spin-up and -down states, respectively (c.f. Eq.~(\ref{varoriginal})), are eigenstates of $\hat{\sigma}_{\rm bath}^{z}$ as follows: 
\eqn{\label{vargs1}
\hat{\sigma}_{\rm bath}^{z}|\Psi_{\pm}\rangle=\mp|\Psi_{\pm}\rangle.
} 
In the deep AFM phase with $\sigma_{\rm tot}^{z}=0$, one can numerically verify that the projected states $|\Psi_{\rm \pm}\rangle$ satisfy the following relations: 
\eqn{\label{vargs2}
\left\vert\left\vert \left\vert \Psi _{\pm}\right\rangle \right\vert\right\vert =1/2,\;\;\left\langle \Psi_{-}\right\vert \hat{\sigma}_{\mathrm{bath}}^{+}\left\vert \Psi_{+}\right\rangle=-1/2.
}
This automatically results in the vanishing of the impurity magnetization $\langle\hat{\sigma}_{\rm imp}^{z}\rangle=0$ and underscores that the variational ground state is indeed the singlet state between the impurity and the bath spins:
\eqn{\label{AFMvar}
|\Psi_{\rm AFM}\rangle=\frac{1}{\sqrt{2}}\left(|\!\!\uparrow\rangle_{\rm imp}|\Psi_{\downarrow}\rangle-|\!\!\downarrow\rangle_{\rm imp}|\Psi_{\uparrow}\rangle\right),
} 
where we use Eq.~(\ref{varoriginal}) and denote the normalized spin-1/2 bath states as $|\Psi_{\downarrow}\rangle=\sqrt{2}|\Psi_{+}\rangle$ and $|\Psi_{\uparrow}\rangle=-\sqrt{2}|\Psi_{-}\rangle$.

 It is worthwhile to mention that, if we choose $|\Psi_{\rm b}\rangle$ in Eq.~(\ref{varoriginal}) as a single-particle excitation on top of the Fermi sea, Eq.~(\ref{AFMvar}) reproduces the variational ansatz that has been originally suggested by Yosida  \cite{KY66} and then later generalized to the Anderson model  \cite{VCM76} and resonant-state approach  \cite{BG07}. This variational state has been recently revisited \cite{YC17}  and shown to contain the majority of the entanglement in the ground state of the Kondo model, indicating the ability of our variational states to encode the most significant part of the impurity-bath entanglement. Yet, we emphasize that our variational ansatz goes beyond such a simple ansatz since we consider the general Gaussian state that takes into account all the correlations between two fermionic operators. Such a flexibility beyond single-particle excitations becomes crucially important when we do quantitative analyses on the Kondo physics as we will demonstrate later.

 \begin{figure*}[t]
\includegraphics[width=170mm]{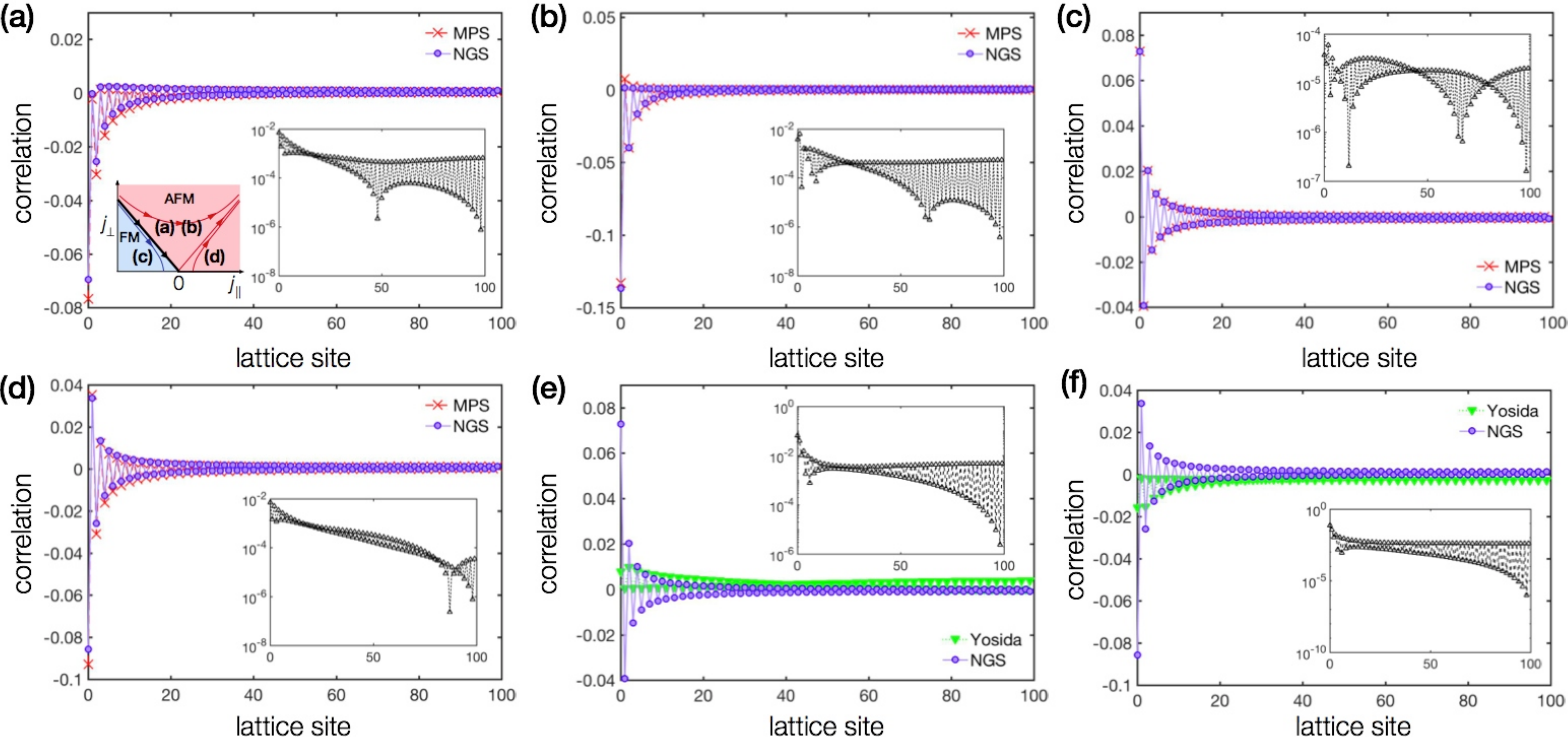} 
\caption{\label{fig_mps1}
Comparisons of ground-state impurity-bath spin correlations $\chi^{z}_{l}=\langle\hat{\sigma}^{z}_{\rm imp}\hat{\sigma}_{l}^{z}\rangle/4$ calculated from our non-Gaussian variational state (NGS, blue circles) and (a-d)  the matrix-product state (MPS, red crosses), and (e,f) the Yosida ansatz (green triangulars). The inset panels show the absolute values of the difference between the correlations calculated from each method.  The dimensionless Kondo couplings $(j_{\parallel},j_{\perp})$ are set to be (a) $(-0.1,0.4)$, (b) (0.1,0.4), (c,e) (-0.4,0.1) and (d,f) (0.4,0.1) as indicated in the left inset panel of (a). System size is $L=100$ and the bond dimension of MPS is $D=260$. 
}
\end{figure*}

In practice, we also monitor the formation of the Kondo-singlet formation by testing the sum rule \cite{BL07} of the impurity-bath spin correlation $\chi_l=\langle\hat{\boldsymbol \sigma}_{\rm imp}\cdot\hat{\boldsymbol \sigma}_{l}\rangle/4$:
\eqn{
\sum_{l=0}^{L}\chi_{l}=\frac{1}{8}\langle\hat{\boldsymbol\sigma}_{\rm tot}^{2}-\hat{\boldsymbol\sigma}_{\rm imp}^{2}-\hat{\boldsymbol\sigma}_{\rm bath}^{2}\rangle=-\frac{3}{4}.
}
In the discussions below, we checked that this sum rule has been satisfied with an error below $0.5\%$ in AFM regime with $J_\parallel>0$.  These observations clearly show that our variational states successfully capture the most important feature of Kondo physics in an efficient manner, i.e., with a number of variational parameters growing only quadratically with the system size $L$.

 \subsection{Comparisons with matrix-product states and Yosida ansatz}
\subsubsection{Matrix-product states}
To further test the accuracy of our approach, we compare our variational results  with those obtained using a MPS ansatz. For the sake of completeness, here we describe the MPS-based method applied to the Kondo model. 
The MPS ansatz for a generic $N$-body quantum system takes the form
\eqn{
|\Psi_{D}\rangle\!=\!\sum_{\{i_k\}} \!\mathrm{Tr}\!\left(A[0]^{i_0} \!A[1]^{i_1}\!\cdots \!A[N-1]^{i_{N\!-\!1}}\right)\!|i_0 i_1\cdots i_{N-1}\rangle,\nonumber\\
}
where for each site $k$, $\{|i_k\rangle\}$ represents a finite basis of the corresponding Hilbert space, and $A[k]^{i_k}$ is a 
$D\times D$ matrix labelled by an index $i_k$. We note that, with open boundary conditions, the first and last matrices reduce to $D$-dimensional vectors.
The parameter $D$ is known as bond dimension, and determines the number of variational parameters in the ansatz.
The MPS ansatz can be used to approximate ground states and low lying excitations of strongly correlated quantum many-body systems. It can also be employed to simulate real-time dynamics and lies at the basis of the DMRG method  \cite{white92dmrg,verstraete04dmrg,vidal2003mps,SU11,Vidal2004,daley04adapt,FV08}.

In order to find a MPS approximation to the ground state of Hamiltonian~(\ref{anisotropicK}), we express the Hamiltonian as a matrix product operator  (MPO)  \cite{pirvu10mpo}.
For the sake of convenience, we actually work in terms of  the bath modes in Eq.~(\ref{compbasis}) and map them to spins using a Jordan-Wigner transformation
such that $\hat{\Psi}_{l \uparrow}=\Pi_{k<l}( \hat{\sigma}_{2k}^z\hat{\sigma}_{2k+1}^z) \hat{\sigma}_{2l}^-$ and 
$\hat{\Psi}_{l \downarrow}=\Pi_{k<l}( \hat{\sigma}_{2k}^z\hat{\sigma}_{2k+1}^z) \hat{\sigma}_{2l}^z \hat{\sigma}_{2l+1}^-$.
A MPS approximation to the ground state is found by variational minimization of the energy over the family of MPS with fixed bond dimension $D$, 
\begin{equation}
|\Psi_0\rangle=\mathrm{argmin} \frac{\langle\Psi_D| H |\Psi_D\rangle}{\langle\Psi_D| \Psi_D\rangle}.
\end{equation}
To solve the minimization, an alternating least squares strategy is applied, where all tensors but one are fixed, and the problem is transformed in the optimization for a single local tensor.
Once the local problem is solved, the optimization is repeated with respect to the next site in the chain and so on 
until the end of the chain is reached. The sweeping is iterated back and forth over the chain until the energy of the
ground state is converged to the desired precision level
(see e.g., Refs. \cite{SU11,FV08} for details of technical details).
The algorithm produces a MPS candidate for the ground state, and local expectation values and correlations
can be efficiently evaluated.
To improve the precision, the procedure can be repeated with increasing bond dimension using the previously found state as the initial guess. Also the convergence criterion can be refined for more accurate results. For our comparisons we set a convergence cutoff of $10^{-8}$-$10^{-6}$.

We can also use the MPS ansatz for a time-evolved state after the quench protocol considered in this paper (see explanations in the next subsection). Although the initial state, i.e., the Fermi sea of the bath (c.f. Eq.~(\ref{initialKondo}) below), is not an exact MPS, we can find a good approximation to it by running the ground state algorithm described above in the absence of the impurity.
Then, the full initial state obtained by a tensor product with the polarized impurity is evolved with the full Hamiltonian.
To this end, the evolution operator is approximated by a Suzuki-Trotter decomposition  \cite{trotter59,Suzuki1985} as a product of small discrete time steps, each of which can be written as a product of MPO  \cite{Vidal2004,Verstraete2004a,pirvu10mpo}. The action of the latter on the  state is then approximated by a new MPS that represents the evolved state.
This is achieved again by an alternating least squares algorithm and minimizing the distance between the MPS and the result of each evolution step (c.f. Refs.  \cite{SU11,FV08}).
In general, real-time evolutions may quickly increase the entanglement in the state, and to maintain an accurate MPS approximation the bond dimension of the ansatz will need
to grow with time. In order to keep track of the accuracy of the simulations, we check convergence of the results at different times when we repeat the simulations with increasing bond dimension.

 \begin{figure}[t]
\includegraphics[width=80mm]{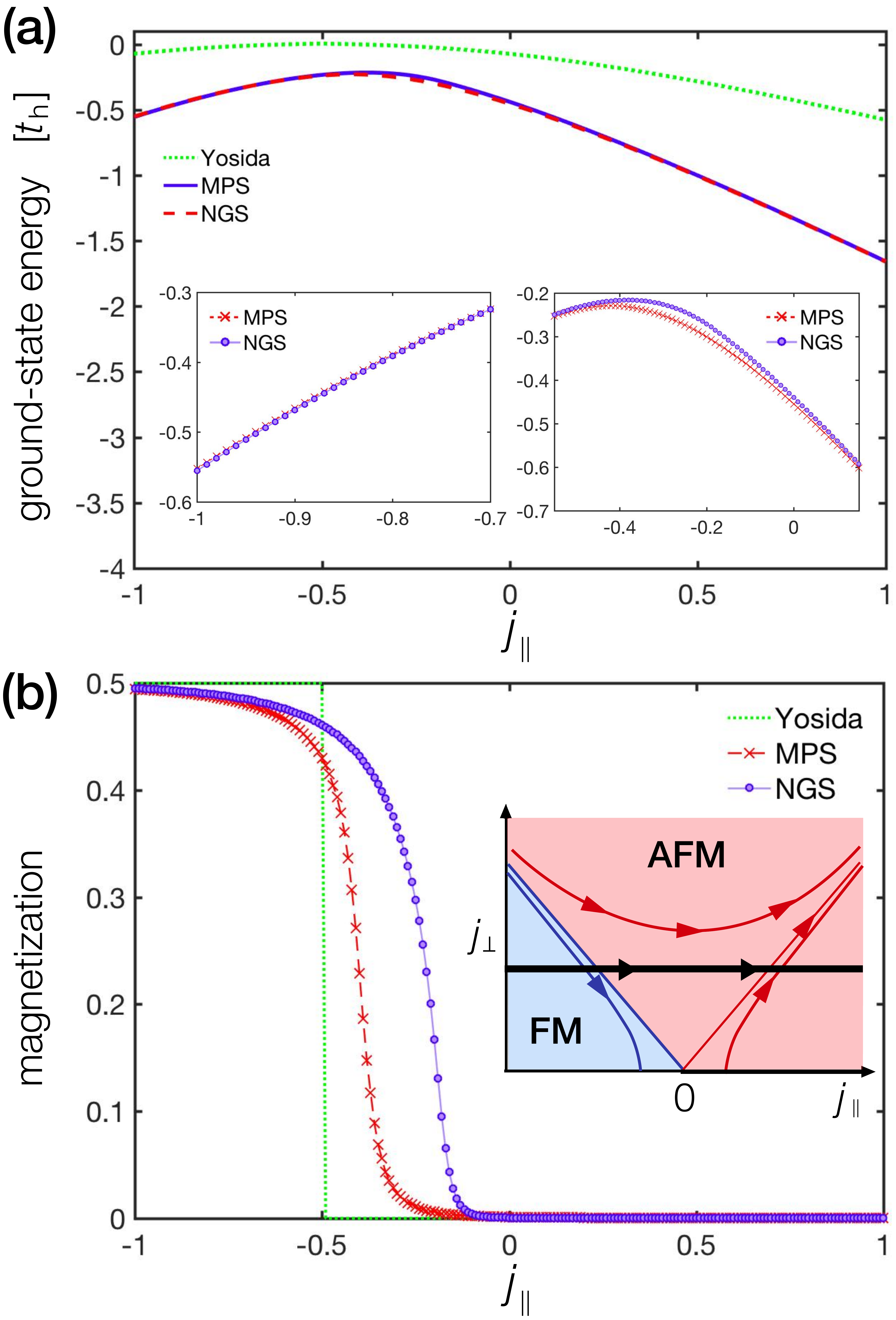} 
\caption{\label{fig_mps2}
Comparison of (a) the ground-state energy and (b) impurity magnetization 
across the phase boundary of the anisotropic Kondo model calculated from the Yosida ansatz (green dotted line), the matrix-product state (MPS, red dashed line with crosses) and our non-Gaussian variational state (NGS, blue solid line with circles). In (a), we plot the variational energies $E_{\rm var}-E_{f}$ relative to a constant ground-state energy $E_{f}$ of free fermions on the lattice. The left and right insets magnify the ground-state energies in the FM phase and close to the phase boundary, respectively.
We choose the dimensionless Kondo coupling $j_{\perp}=0.5$ and vary $j_{\parallel}$ from $-1$ to $1$ as schematically illustrated in the inset of (b). The calculations are done in the sector $\sigma_{\rm tot}^{z}=0$. System size is $L=200$ and the bond dimension of MPS is $D=280$. 
}
\end{figure}

\subsubsection{Yosida ansatz}
To highlight the importance of taking into account contributions beyond a single-particle excitation, we also compare our results to the much simpler ansatz originally suggested by Kei Yosida \cite{KY66}. This ansatz assumes a single-particle excitation on top of the half-filled Fermi sea $|{\rm FS}\rangle$. To deal with the anisotropic Kondo model, we consider the following ansatz:
\begin{widetext}
\eqn{\label{yosidaansatz}
|\Psi_{\rm Yosida}\rangle=\begin{cases}
\frac{1}{\sqrt{2}}\sum_{n>n_{\rm F}}d_{n}\left(|\!\uparrow\rangle_{\rm imp}\hat{c}^{\dagger}_{n\downarrow}-|\!\downarrow\rangle_{\rm imp}\hat{c}_{n\uparrow}^{\dagger}\right)|{\rm FS}\rangle & (-J_{\parallel}\leq J_{\perp}),\\
\sum_{n>n_{\rm F}}d_{n}|\!\uparrow\rangle_{\rm imp}\hat{c}^{\dagger}_{n\uparrow}|{\rm FS}\rangle & (-J_{\parallel}> J_{\perp}),
\end{cases}
}
\end{widetext}
where $n$ denotes the bath mode in the energy basis and the summation over $n>n_{\rm F}$ indicates the contributions from the bath modes above the Fermi surface. The former (latter) ansatz in Eq.~(\ref{yosidaansatz}) approximates the singlet (triplet) state in the AFM (FM) regime. We minimize the mean energy $\langle\hat{H}\rangle_{\rm Yosida}$ with respect to amplitudes $\{d_{n}\}$ and obtain the following variational equation:
\eqn{\label{varyosida}
\epsilon_{n}d_{n}-\frac{1}{4}J_{\rm eff}\psi_{0n}^{*}\sum_{m>n_{\rm F}}\psi_{0m}d_{m}=E_{\rm GS}d_{n},
}
where $\epsilon_{n}$ denotes a bath energy, $\psi_{ln}$'s are expansion coefficients in terms of the lattice basis, $\hat{c}_{l}=\sum_{n}\psi_{ln}\hat{c}_{n}$, and $E_{\rm GS}$ is the variational ground-state energy. We define $J_{\rm eff}$ as the effective Kondo coupling as follows:
\eqn{
J_{\rm eff}=
\begin{cases}
2J_{\perp}+J_{\parallel}&(-J_{\parallel}\leq J_{\perp}),\\
-J_{\parallel}&(-J_{\parallel}> J_{\perp}).
\end{cases}
}
We solve the eigenvalue equation~(\ref{varyosida}) and determine $E_{\rm GS}$ and the wavefunction $d_{n}$ from which the impurity-bath correlation can be calculated. It is worthwhile to mention that the Yosida ansatz is naturally included in our family of variational states. For instance, in the AFM regime the singlet entanglement can be decoupled in the transformed frame as
\eqn{
\hat{U}^{-1}|\Psi_{\rm Yosida}\rangle=|+_{x}\rangle\sum_{n>n_{\rm F}}\frac{d_{n}}{\sqrt{2}}\left(\hat{c}_{n\downarrow}^{\dagger}-\hat{c}_{n\uparrow}^{\dagger}\right)|{\rm FS}\rangle.
}
Here, the bath wavefunction in the transformed frame just corresponds to a single-particle excitation on top of the Fermi sea, which is a very special subclass of the whole fermionic Gaussian states.

\subsubsection{Benchmark results}

We plot in Fig.~\ref{fig_mps1} the ground-state impurity-bath spin correlations $\chi^{z}_{l}=\langle\hat{\sigma}^{z}_{\rm imp}\hat{\sigma}_{l}^{z}\rangle/4$ of the anisotropic Kondo model in four different regimes of the phase diagram (see the left inset of Fig.~\ref{fig_mps1}(a)). We use the dimensionless Kondo couplings $j_{\parallel,\perp}=\rho_{F}J_{\parallel,\perp}$ with $\rho_{F}=1/(2\pi t_{\rm h})$ being the density of states at the Fermi energy.
Our variational results not only correctly reproduce the formations of the singlet (triplet) pair between the impurity and bath spins in the AFM (FM) phase, but they also show quantitative agreement with MPS results with a deviation which is at most a few percent of the value at the impurity site. We find that the agreement is particularly good in the deep FM and AFM regimes (see Figs.~\ref{fig_mps1}(c) and (d)), where the difference is below $1$\%. We note that while the Yosida ansatz qualitatively captures the correct AFM and FM correlations in each regime, it fails to agree with our variational results and MPS ones quantitatively (Figs.~\ref{fig_mps1}(e) and (f)).

We also compare the ground-state energy $E_{\rm var}$ (Fig.~\ref{fig_mps2}(a)) and the corresponding magnetization $\langle\hat{\sigma}_{\rm imp}^{z}\rangle$ (Fig.~\ref{fig_mps2}(b)) across the phase transition line (see the inset of Fig.~\ref{fig_mps2}(b)). In the FM phase, we observe that our variational results agree very well with the MPS results, with an error below $0.5$\%. Remarkably, our ansatz achieves slightly lower energies deep in the phase (the left inset of Fig.~\ref{fig_mps2}(a)). 
Close to the phase boundary, we observe the largest deviation with respect to the MPS ansatz both in energy (the right inset of Fig.~\ref{fig_mps2}(a)) and magnetization (Fig.~\ref{fig_mps2}(b)), with our ansatz showing a residual magnetization close to the transition.
Finally, in the deep AFM phase ($j_\parallel>0$), the ground-state energies calculated from both methods again agree well with an error typically below $0.5$\%. In this regime, the corresponding residual magnetization is $\langle\hat{\sigma}_{\rm imp}^{z}\rangle\simeq O(10^{-4})$ in MPS and  $O(10^{-5})$ in our variational method. We note that the Yosida ansatz gives significantly higher ground-state energies than our results and MPS ones over the whole phase diagram. 
We attribute the small discrepancies between our results and the MPS ones to the finite values of the system size $L$ and the bond dimension $D$. 
A relatively large difference in the AFM phase close to the phase boundary should be attributed to the large entanglement present in this regime. This fact can be inferred from the nonmonotonic RG flows, which show that the Kondo couplings can take very small values corresponding to the generation of the large Kondo cloud during the flows. As our calculations are carried out in the real-space basis, this fact implies that our variational states must encode such a large entanglement, which can be beyond the amount that is generated by the unitary transformation $\hat{U}$. 

The comparisons to the MPS results show the great efficiency of our variational approach. The number of variational parameters in MPS is approximately $4LD^2$ with $D$ being the bond dimension, while that of our variational ansatz is $4L^2$. Since the bond dimension $D$ in the calculations is typically taken to be 200-300, our variational approach can achieve the accuracy comparable to MPS with two or three orders of magnitude fewer variational parameters, and shorter CPU time accordingly, than the corresponding MPS ansatz. This indicates that our variational ansatz successfully represents the ground-state wavefunction of SIM in a very compact way. Meanwhile, the comparison to the Yosida ansatz clearly demonstrates the importance of considering contributions beyond the single-particle excitations. Equivalently, it indicates that one can dramatically improve the variational calculations by just taking into account up to two-fermionic excitations. This significant simplification of the original many-body problems is made possible through the decoupling transformation that can efficiently encode the impurity-bath entanglement.

 \begin{figure}[t]
\includegraphics[width=86mm]{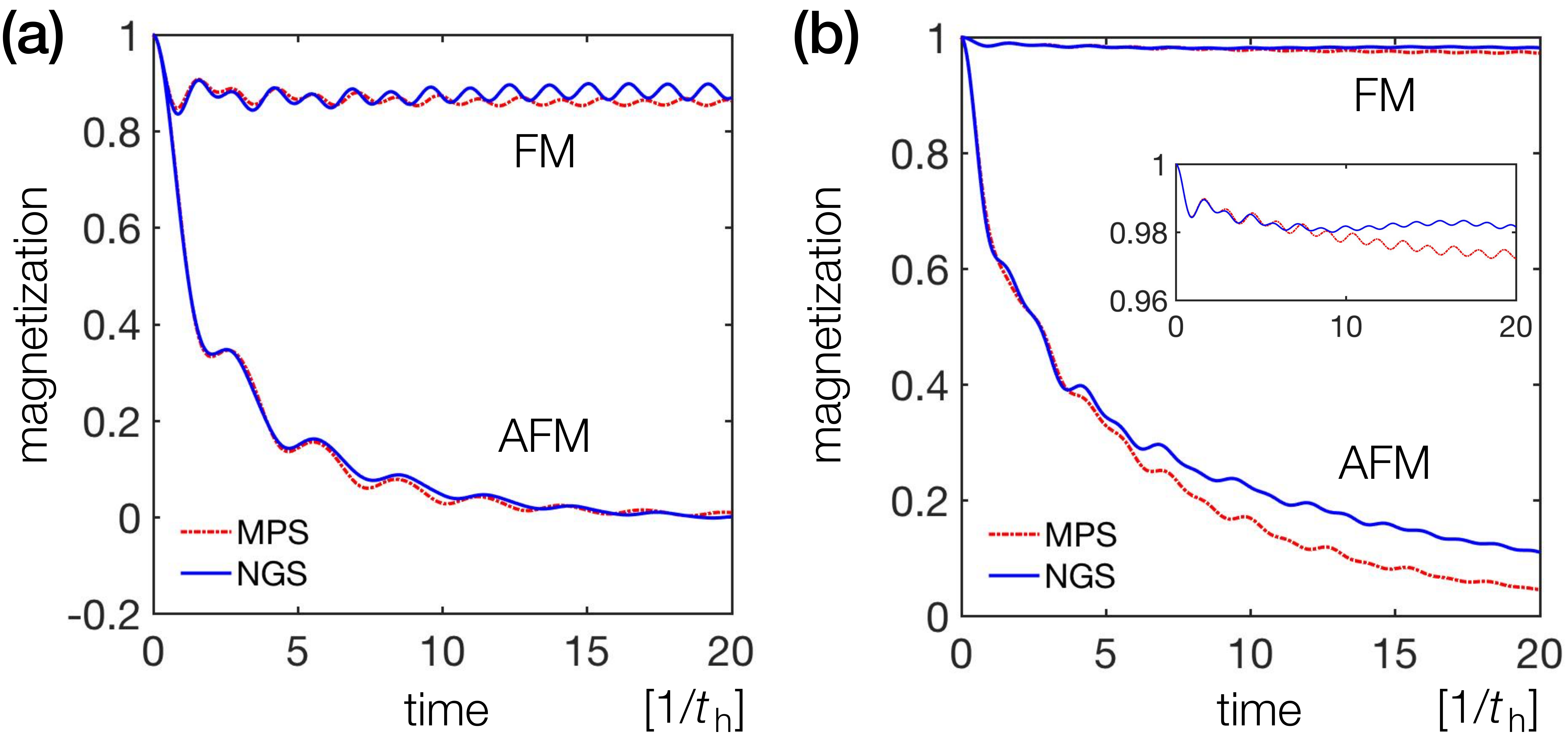} 
\caption{\label{fig_mps_RT}
Comparisons of dynamics of the impurity magnetization calculated from the matrix-product state (MPS, red chain curve) and our non-Gaussian variational state (NGS, blue solid curve) at (a) SU(2)-symmetric points and (b) strongly anisotropic regimes. The dimensionless Kondo couplings are (a) $j_{\parallel}=j_{\perp}=\pm0.35$ in the SU(2)-symmetric antiferromagnetic (AFM) and ferromagnetic (FM) phases, respectively, and (b) $(j_{\parallel},j_\perp)=(0.1,0.4)$ and $(-0.4,0.1)$ in the anisotropic AFM and FM phases, respectively. The inset in (b) magnifies the dynamics in FM phase.  System size is $L=100$ and the bond dimension of the MPS for the different cases varies in $D\in[220,\,260]$, for which we checked that the results are  converged within the time window shown in the plots.
}
\end{figure}

Finally, to complete the benchmark test in the anisotropic Kondo model, we compare the real-time dynamics of the impurity magnetization calculated with a MPS algorithm and our variational method. We consider the initial state
\eqn{\label{initialKondo}
|\Psi(0)\rangle=|\!\uparrow\rangle_{\rm imp}|{\rm FS}\rangle,
}
where $|{\rm FS}\rangle$ is the half-filled Fermi sea of the lead. At time $t=0$, we then drive the system by suddenly coupling the impurity to the lead.

 First, we show the results at the SU(2)-symmetric points of AFM and FM phases (Fig.~\ref{fig_mps_RT}(a)), which are usually of interest in condensed matter systems. The magnetization eventually relaxes to zero in the AFM phase as a result of the Kondo-singlet formation, while it remains nonvanishing and exhibits oscillations in the FM phase. The long-lasting fast oscillation found in the FM phase comes from a high-energy excitation of a fermion from the bottom of the band and its period $2\pi\hbar/{\cal D}$ is characterized by the bandwidth ${\cal D}=4t_{\rm h}$ (note that we choose the unit $\hbar=1$ and $t_{\rm h}=1$ in the plots). Since the impurity spin will be decoupled from the bath degrees of freedom in the low-energy limit in the FM phase (see RG flows in Fig.~\ref{fig_mps2}(b) inset), the oscillation can survive even in the long-time limit.  The MPS and our variational results show quantitative agreement with an error that is at most $O(10^{-2})$. Second, we compare the results in the strongly anisotropic AFM and FM regimes (Fig.~\ref{fig_mps_RT}(b)). In the short and intermediate time regimes, the results obtained using MPS and our variational states exhibit a good agreement. In the long-time regime ($t\gtrsim10$), the two methods show a small discrepancy that is about $O(6\times 10^{-2})$ in the AFM phase and $O(1\times 10^{-2})$ in the FM phase. Yet, they still share the qualitatively same features such as the exponential relaxation in the AFM regime and long-lasting oscillations characterized by the bandwidth in the FM regime.
 
\subsection{Comparison with the Bethe ansatz solution}

\begin{figure}[t]
\includegraphics[width=70mm]{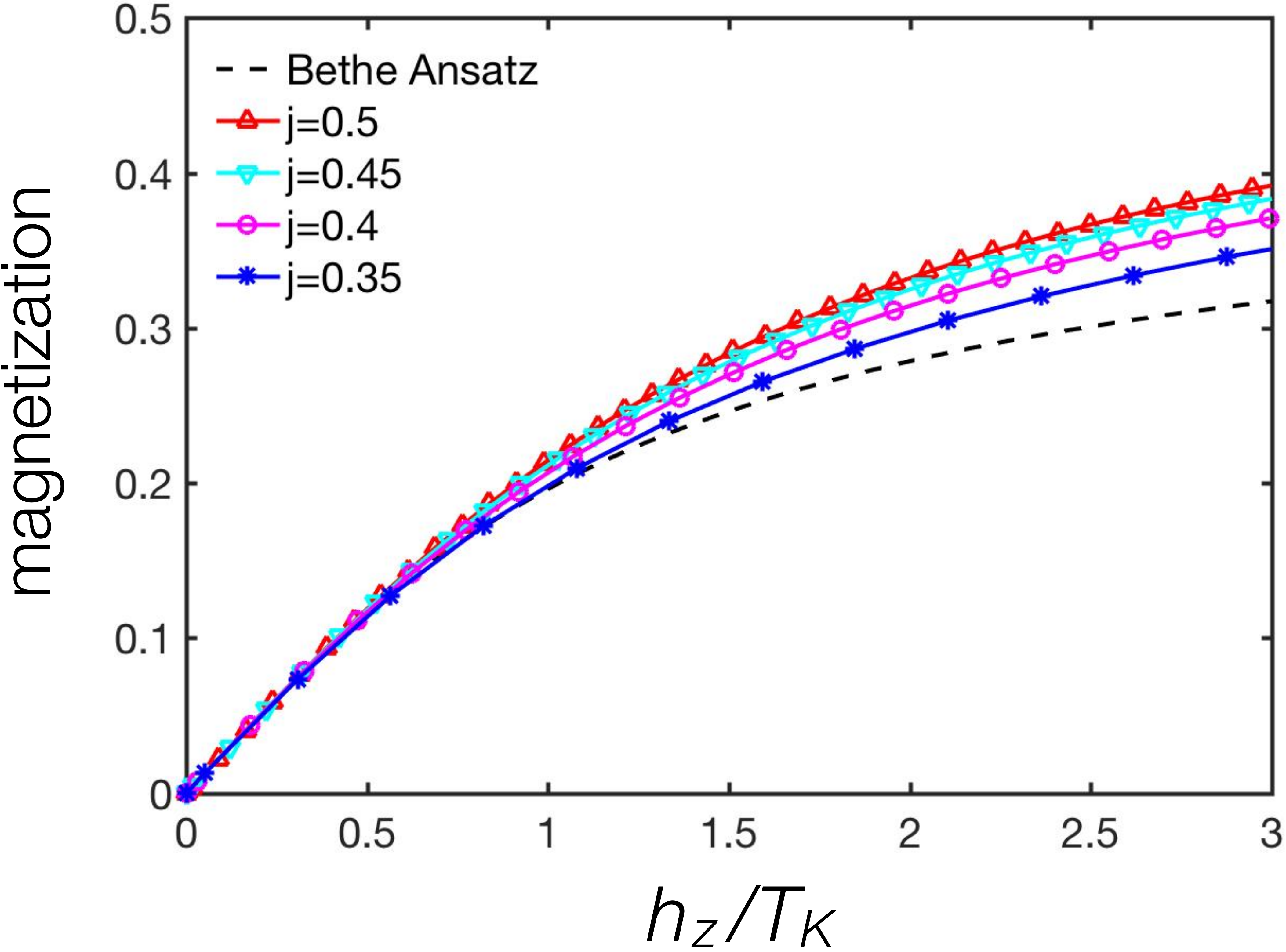} 
\caption{\label{fig_Bethe}
Comparison of the zero-temperature magnetization $m=\langle\hat{\sigma}_{\rm imp}^{z}\rangle/2$ calculated from our approach for different Kondo couplings $j$ with the Bethe ansatz solution (dashed black line), which is exact under the infinite-bandwidth assumption. The deviation at a large $h_z/T_{\rm K}$ from the Bethe ansatz solution originates from a finite bandwidth. The deviations begin at smaller $h_z/T_{\rm K}$ for a larger coupling $j$ since the latter leads to a larger Kondo temperature $T_{\rm K}$. The Kondo temperature is extracted from the magnetic susceptibility at the zero field $\chi=1/(4T_{\rm K})$.
}
\end{figure}

As a further test of the validity of the present approach, we compare our approach to the exact solution obtained via the Bethe ansatz with the infinite-bandwidth assumption \cite{PBW80,AND80,ANL81,NK81,AN83,SP89}. When all the physical parameters are smaller than the bandwidth ${\cal D}=4t_{\rm h}$, our approach should reproduce the universal prediction from the Bethe ansatz. To be specific, we calculate the zero-temperature magnetization $m=\langle\hat{\sigma}_{\rm imp}^{z}\rangle/2$ under a finite magnetic field $h_z$ and compare the results to the one predicted from the Bethe ansatz (BA) solution \cite{ANL81}:
\begin{widetext}
\eqn{
m_{\rm BA}=\begin{cases}
\frac{1}{4\sqrt{\pi^3}}\sum_{k=0}^{\infty}\frac{(-1)^k}{k!}\left(\frac{\pi}{2}\right)^{2k+1}\left(\frac{k+1/2}{2\pi}\right)^{k-\frac{1}{2}}\left(\frac{h_z}{T_{\rm K}}\right)^{2k+1}& (h_z/T_{\rm K}\leq\sqrt{8/(\pi e)}),\\
\frac{1}{2}\Biggl\{1-\frac{1}{\pi^{3/2}}\int_{0}^{\infty}dt\frac{\sin(\pi t)}{t}\left[\frac{8}{\pi e}\left(\frac{T_{\rm K}}{h_z}\right)^2\right]e^{-t(\ln t-1)}\Gamma\left(t+\frac{1}{2}\right)\Biggr\}& (h_z/T_{\rm K}>\sqrt{8/(\pi e)}).
\end{cases}
}
\end{widetext}
Here, we note that the Kondo temperature $T_{\rm K}$ is determined from the magnetic susceptibility $\chi\equiv\partial m/\partial h_{z}=1/(4T_{\rm K})$ at the zero field $h_z=0$. Figure~\ref{fig_Bethe} shows the comparison between the magnetization obtained from our variational method for different Kondo couplings $j$ and the Bethe ansatz solution $m_{\rm BA}$ (dashed black line). The numerical results agree with $m_{\rm BA}$ with at most a few percent deviation at small magnetic field $h_z/T_{\rm K}\lesssim 1$. As we increase the magnetic field $h_z/T_{\rm K}$, the deviation from $m_{\rm BA}$ becomes more significant. In particular, the deviation starts at smaller $h_z/T_{\rm K}$ for a larger Kondo coupling $j$. These features should be attributed to a finite-bandwidth effect intrinsic to the lattice model. Indeed, at the largest value of $h_z/T_{\rm K}\simeq 3$ the value of magnetic field $h_z$ itself can be an order of the bandwidth ${\cal D}=4t_{\rm h}$. The larger Kondo coupling $j$ corresponds to the larger Kondo temperature $T_{\rm K}$, thus leading to the smaller threshold value of $h_z/T_{\rm K}$ at which the finite-bandwidth effect begins to take place.

\section{Application to the two-lead Kondo model}\label{secKondotwo}

\begin{figure}[b]
\includegraphics[width=56mm]{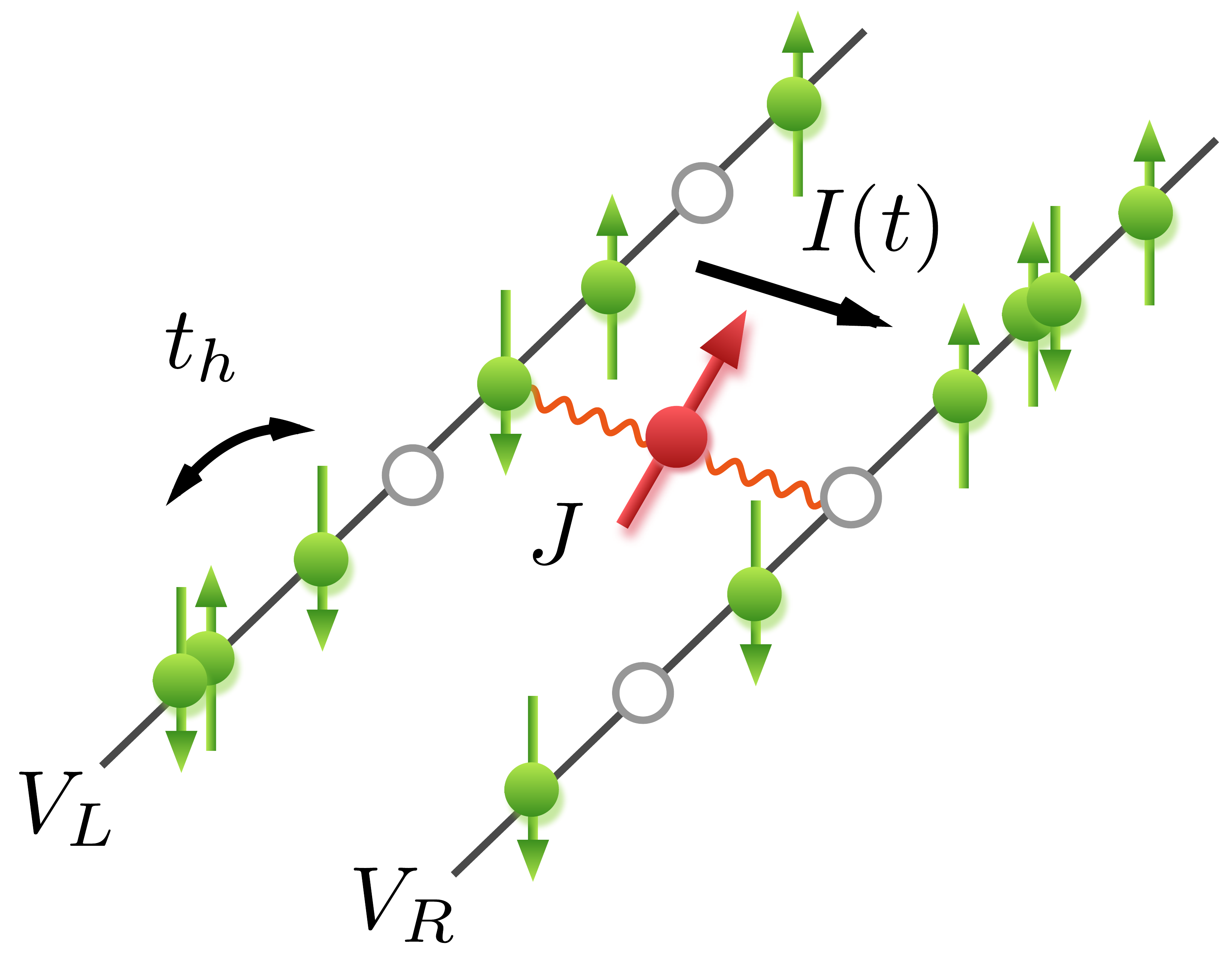} 
\caption{\label{fig_trans_s}
Schematic figure of the two-lead Kondo model. The localized spin-1/2 impurity is coupled with the centers of the left (L) and right (R) leads via an isotropic Kondo coupling $J$. The bath fermions can move within each lead with a hopping $t_{\rm h}$. The bias potentials $V_{\rm L,R}$ are applied to the left and right leads. The time evolutions of the current $I(t)$  between the two leads, the impurity magnetization $\langle\hat{\sigma}^{z}_{\rm imp}(t)\rangle$, and spatially resolved bath properties can be efficiently  calculated by our variational method.
}
\end{figure}

\begin{figure*}[t]
\includegraphics[width=170mm]{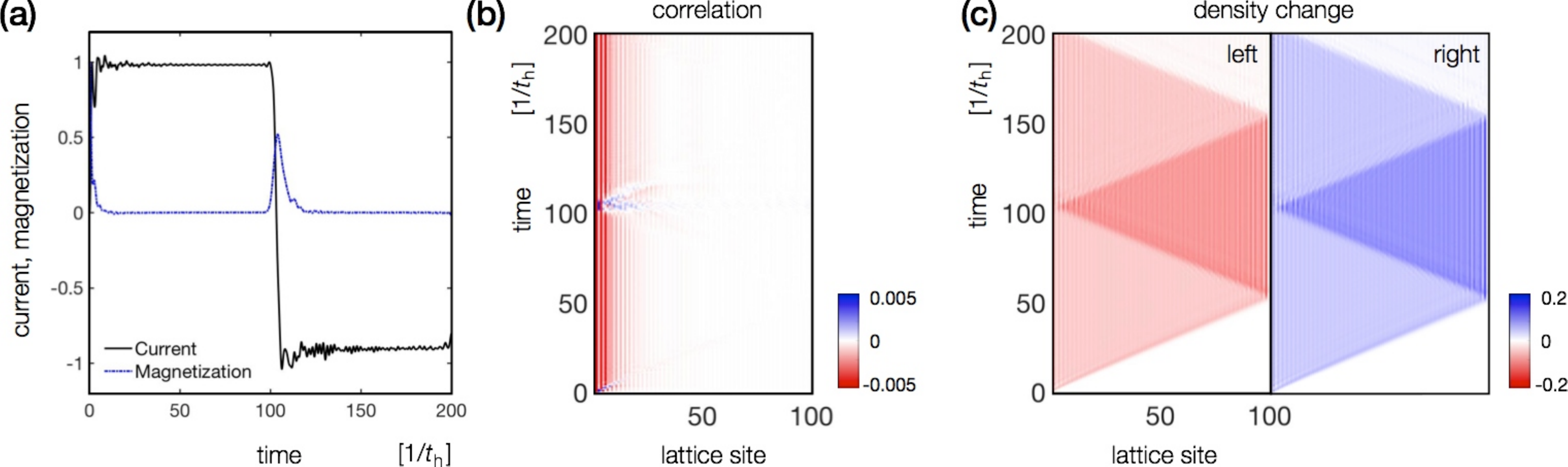} 
\caption{\label{fig_trans1}
Out-of-equilibrium dynamics in the two-lead Kondo model. (a) The time evolutions of the current $I(t)$ (black solid curve) and the impurity magnetization $\langle\hat{\sigma}_{\rm imp}^{z}(t)\rangle$ (blue chain curve). The current is plotted in unit of $et_{\rm h}/h$. After the quench of the chemical potentials and the Kondo coupling at $t=0$, the current quickly reaches a steady value and exhibits the plateau, while the impurity magnetization quickly decays to zero, indicating the Kondo-singlet formation. The current suddenly flips its sign in the middle of the time evolution at which the density wave returns to the impurity site after reflecting at the end of each lead. In this regime, the magnetization passingly takes a large non-zero value. (b) and (c) Spatiotemporal dynamics of (b) the impurity-bath spin correlation function $\chi^{z}_{l}(t)$ and (c) the relative changes of density in the left and right lead. At $t\simeq 50/t_{\rm h}$, the density waves reach the ends of each lead. They return to the impurity site $l=0$ at $t\simeq 100/t_{\rm h}$, where the spin correlations passingly become  ferromagnetic. When the density waves return to the impurity site at $t\simeq 200/t_{\rm h}$ after the second reflection at the lead ends at $t\simeq 150/t_{\rm h}$, the density reproduces the initial homogeneous profile. The parameters are $j=0.4$, $V_{\rm L}=-V_{\rm R}=0.25t_{\rm h}$. System size if $L=100$ for each lead.
}
\end{figure*}

\subsection{Model}

We next apply our approach to investigate out-of-equilibrium dynamics and transport properties in the two-lead Kondo model  \cite{KA00}  (Fig.~\ref{fig_trans_s}), where the localized spin impurity is coupled to the centers of the left and right leads via the isotropic Kondo coupling. The Hamiltonian is

\eqn{\label{two-leads}
\hat{H}_{\rm two}&=&\sum_{l\eta}\biggl[-t_{\rm h}\bigl(\hat{c}^{\dagger}_{l\eta\alpha}\hat{c}_{l+1\eta\alpha}\!+\!{\rm h.c.}\bigr)+eV_{\eta}\,\hat{c}^{\dagger}_{l\eta\alpha}\hat{c}_{l\eta\alpha}\biggr]\nonumber\\
&&+\frac{J}{4}\sum_{\eta\eta'}\hat{\boldsymbol{\sigma}}_{\rm imp}\cdot\hat{c}_{0\eta\alpha}^{\dagger}\boldsymbol{\sigma}_{\alpha\beta}\hat{c}_{0\eta'\beta}-\frac{h_z}{2}\hat{\sigma}_{\rm imp}^{z},
}
where $\hat{c}_{l\eta\alpha}^{\dagger}$ ($\hat{c}_{l\eta\alpha}$) creates (annihilates) a fermion with position $l$ and spin $\alpha$ on the left ($\eta={\rm L}$) or right ($\eta={\rm R}$) lead, the hopping $t_{\rm h}=1$ sets the energy unit, $J$ is an isotropic Kondo coupling strength, and $eV_{\eta}$  are chemical potentials for each lead. The spin indices $\alpha,\beta$ are understood to be contracted as usual.

In the same manner as in the single-lead case in the previous section, only the following symmetric modes are coupled to the impurity:
\eqn{\label{compbasis2}
\hat{\Psi}_{0\eta,\alpha}=\hat{c}_{0\eta\alpha},\;\;\;
\hat{\Psi}_{l\eta,\alpha}=\frac{1}{\sqrt{2}}\left(\hat{c}_{l\alpha\eta}+\hat{c}_{-l\eta\alpha}\right)
}
with $l=1,2,\ldots,L$.
Our formalism in Sec.~\ref{secformalism} can be applied by identifying the bath Hamiltonian $h_{lm}$ as the following two-lead matrix $h_{2}$:
\eqn{\label{two-lead1}
h_2 &=&\left(
\begin{array}{cc}
h_1+eV_{\mathrm{L}}\mathrm{I}_{L+1} & 0 \\
0 & h_1+eV_{R}\mathrm{I}_{L+1}%
\end{array}%
\right),
}
where we recall that $h_1$ is the single-lead hopping matrix given in Eq.~(\ref{hopping1}). The coupling matrix $g_{lm}^{\gamma}$ is given by the local two-lead Kondo coupling:
\eqn{\label{two-lead2}
g^{\gamma } &=&J\left(
\begin{array}{cc}
1 & 1 \\
1 & 1%
\end{array}%
\right) \otimes \mathrm{diag}_{L+1}(1,0,...,0),
}
where ${\rm diag}_{d}(v)$ denotes a $d\!\times\! d$ diagonal matrix with elements $v$.
Using Eqs.~(\ref{two-lead1}) and (\ref{two-lead2}), together with the general expression of the functional derivative (\ref{hm}), we solve the variational real-time evolution~(\ref{realvar}) to study the out-of-equilibrium dynamics and transport properties.

\subsection{Quench dynamics}
To analyze transport phenomena, we prepare the initial state 
\eqn{
|\Psi(0)\rangle=|\uparrow\rangle_{\rm imp}|{\rm FS}\rangle_{\rm L}|{\rm FS}\rangle_{\rm R},
}
where $|{\rm FS}\rangle_{\rm L,R}$ are the half-filled Fermi sea of each lead and, at time $t=0$, we drive the system to out of equilibrium by suddenly coupling the impurity and applying bias potentials $V_{\rm L}=V/2$ and $V_{\rm R}=-V/2$. We then calculate the real-time evolution by integrating  Eq.~(\ref{realvar}). Without loss of generality, we choose a bias $V>0$ so that the
current initially takes a positive value (particles flow from the the left to right lead).

\begin{figure}[b]
\includegraphics[width=71mm]{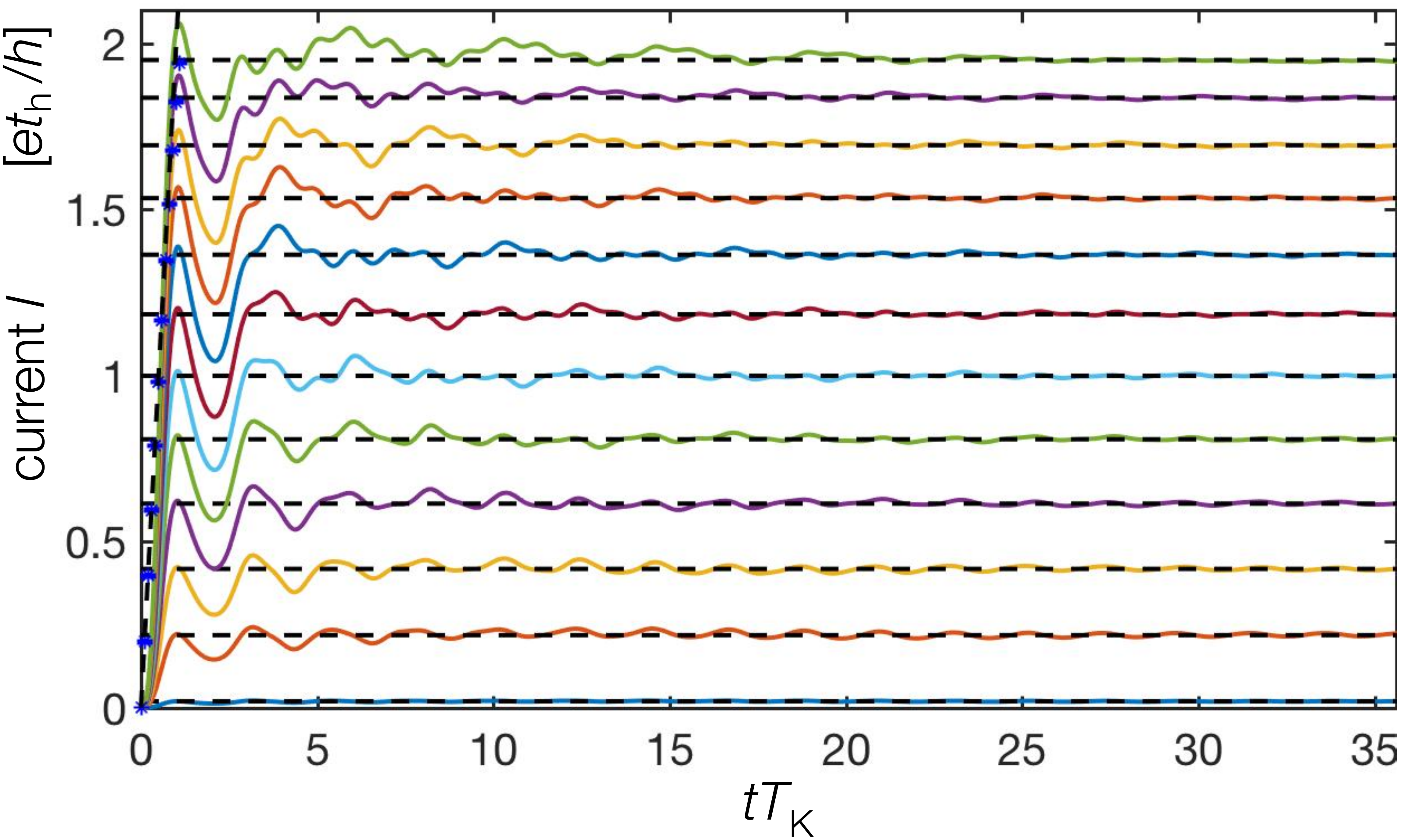} 
\caption{\label{fig_trans_p}
Time evolutions of the current $I(t)$ after the quench for various bias potentials $V$. After transient dynamics, all the currents reach their steady values that are shown as the black dashed lines. The current is plotted in unit of $et_{\rm h}/h$. The dimensionless Kondo coupling is $j=0.4$ and we set a bias potential $V=V_0+\Delta V/2$ with $\Delta V=0.01t_{\rm h}$ and vary $V_0=0,0.1,\cdots,1.1t_{\rm h}$ from the bottom to top. System size is $L=100$ for each lead.
}
\end{figure}

\begin{figure}[b]
\includegraphics[width=86mm]{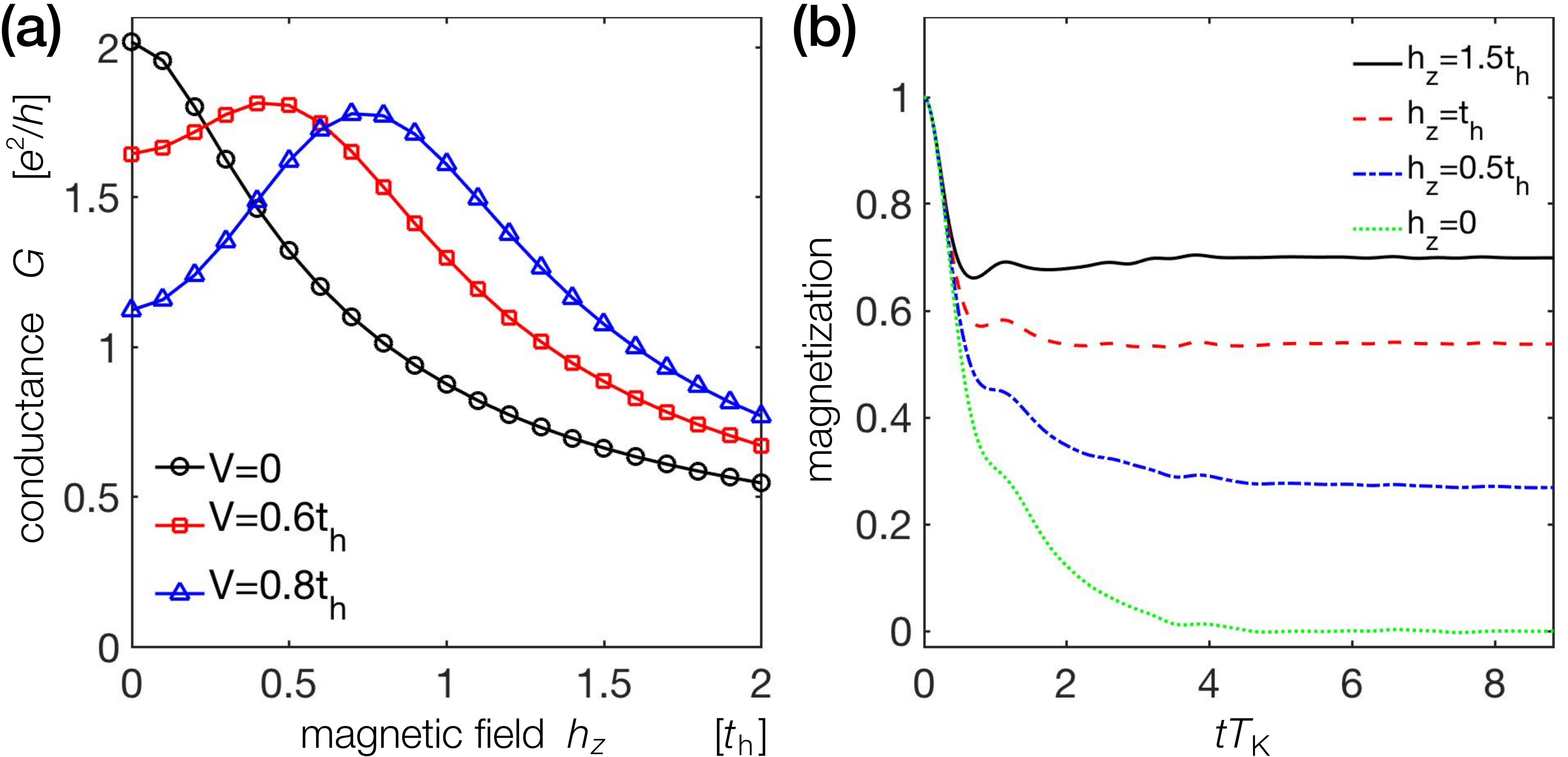} 
\caption{\label{fig_trans_2}
(a) The differential conductance $G$ (in unit of $e^2/h$) is plotted against a magnetic field $h_z$ for different bias potentials $V$. (b) Time evolution of the impurity magnetization $\langle\hat{\sigma}_{\rm imp}^{z}(t)\rangle$ after the quench at finite bias $V$ and magnetic field $h_z$. System size is $L=100$ for each lead and we use $j=0.35$,  $V=0.8t_{\rm h}$ in (b).
}
\end{figure}

The black solid curve in Fig.~\ref{fig_trans1}(a) shows the time evolution of  the current $I(t)$ flowing between the two leads:
\eqn{
I(t) & &=i\frac{eJ}{4\hbar}\left[\langle \hat{\boldsymbol{\sigma}}_{{\rm imp}}\cdot\hat{c}_{0{\rm L}\alpha}^{\dagger}\boldsymbol{\sigma}_{\alpha\beta}\hat{c}_{0{\rm R}\beta}\rangle -{\rm h.c.}\right]\nonumber\\
 && =-\frac{eJ}{2\hbar}{\rm Im}\biggl[\sigma_{{\rm imp}}^{x}\sigma_{\alpha\beta}^{x}(\Gamma_{f})_{0{\rm L}\alpha,0{\rm R}\beta}\nonumber\\
&& \;\;\;\;\;\; \;\;\;\;\;\; \;\;\;\;+(-i\sigma^{y}+\sigma_{{\rm imp}}^{x}\sigma^{z})_{\alpha\beta}(\Gamma_{f}^{{\rm P}})_{0{\rm L}\alpha,0{\rm R}\beta}\biggr],
}
where we recall that $\langle\cdots\rangle$ is an expectation value with respect to wavefunction in the original frame, and we put back $\hbar$ when the current and conductance are under consideration. After a short transient dynamics, the current quickly reaches a steady value and forms a plateau in the time evolution. The relaxation time here is characterized by the time scale of the Kondo-singlet formation, which is roughly equal to the decaying time of the magnetization (blue chain curve in Fig.~\ref{fig_trans1}(a)). Figures~\ref{fig_trans_p} and \ref{fig_trans_2}(b) demonstrate that these time scales are  characterized by the inverse of the Kondo temperature $1/T_{\rm K}$ with $T_{\rm K}$ being extracted from the magnetic susceptibility at the zero field as we discussed before. The stability of the reached steady state can be checked by adding a small perturbation on the dynamics and observing the response of physical observables; if they return back to the original steady values, we can conclude that the reached state is actually stable.

Figure~\ref{fig_trans1}(b) shows the spatiotemporal development of the impurity-bath spin correlation $\chi^{z}_{l}(t)$, which clearly indicates the formation of the Kondo cloud. Figure~\ref{fig_trans1}(c) shows the corresponding spatiotemporal dynamics of the changes in density relative to the initial value. After the quench, density waves propagate ballistically and eventually reach the ends of leads. Then, the reflected waves come back to the impurity site at the middle of the time evolution. This associates with the sudden flip of the current and recurrence of the magnetization (Fig.~\ref{fig_trans1}(a)). Also, the spin correlation passingly becomes ferromagnetic at this moment (Fig.~\ref{fig_trans1}(b)). After the second reflections at the ends of leads, the density in both leads turn back to the initial value as shown in Fig.~\ref{fig_trans1}(c). These results demonstrate the ability of our variational method to accurately calculate long-time spatiotemporal dynamics, which is challenging to obtain in the previous approaches. As we consider the global quench protocol, the system acquires an extensive amount of energy relative to the ground state and thus the amount of entanglement can grow significantly in time. This severely limits the applicability of, e.g., DMRG calculations in the long-time regime  \cite{HMF09}. The predicted spatiotemporal dynamics can be readily tested with site-resolved measurements by quantum gas microscopy   \cite{EM11,CM12,FuT15,KAM16,MM15,YA15,RY16}.  

We remark that all the results are converged against a system size $L$ up to the time $L/t_{\rm h}$ at which the density waves come back to the impurity site and the finite-size effects take place. To avoid a finite-size effect, we also recall that the Kondo coupling $J$ should be large enough such that the Kondo length satisfies $\xi_{\rm K}\ll L$ (see also discussions in Ref.~\cite{YA18L}).

\begin{figure}[t]
\includegraphics[width=76mm]{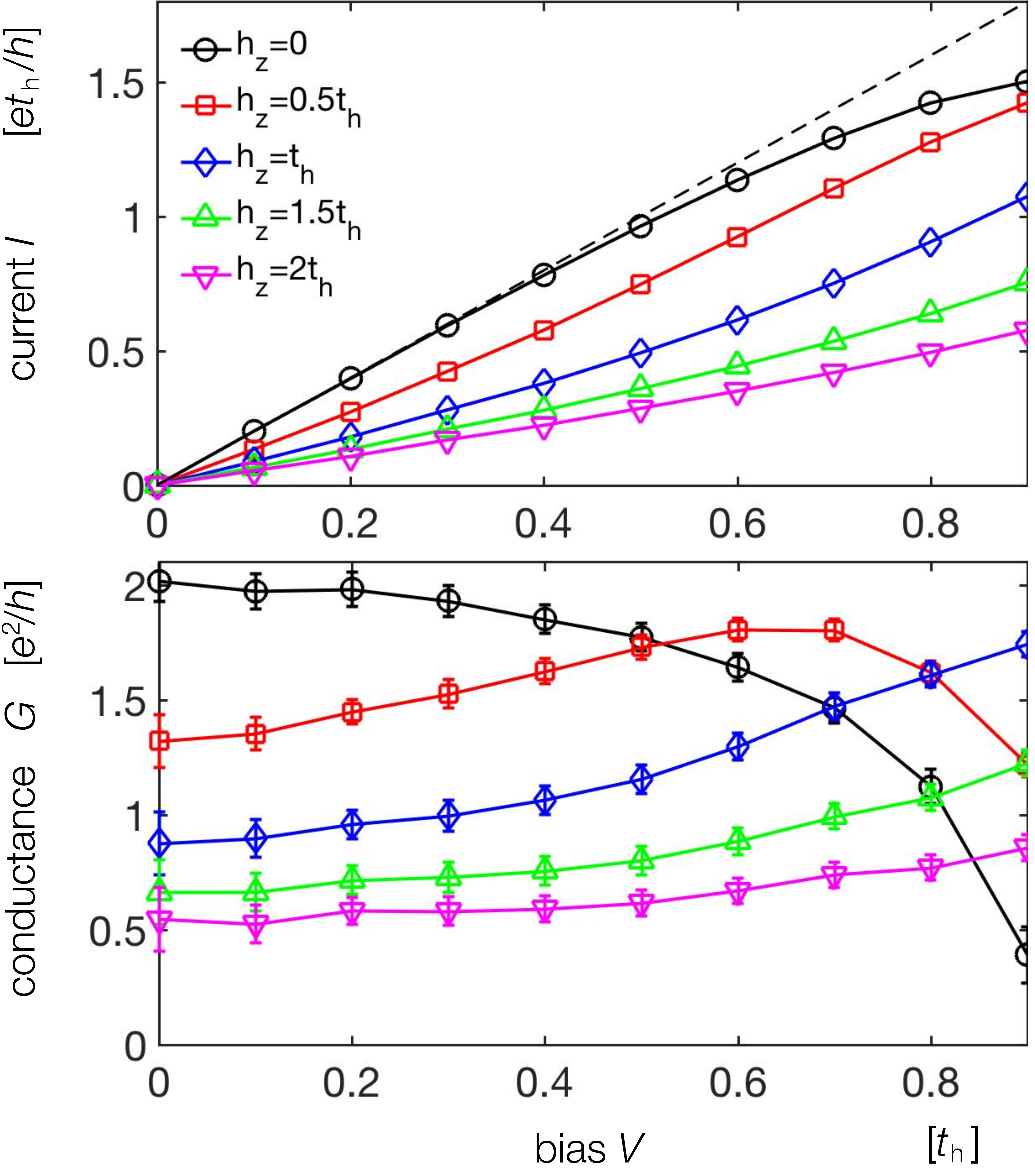} 
\caption{\label{fig_trans_3}
(Top panel) The current-bias characteristics with different strengths of magnetic field $h_z$. The dashed black line indicates the perfect conductance with $I/V=2e^2/h$. The current is plotted in unit of $et_{\rm h}/h$. (Bottom panel) The corresponding plot of the nonlinear differential conductance $G=dI/dV$ against the bias potential $V$ for different magnetic fields $h_z$.
System size is $L=100$ for each lead and we use $j=0.35$ for which the Kondo temperature is obtained as $T_{\rm K}=0.4414t_{\rm h}$ from the magnetic susceptibility $\chi=1/(4T_{\rm K})$. 
}
\end{figure}

\subsection{Conductance at finite bias and magnetic field}
 It has been known that the conductance can exhibit the universal scaling behavior against external parameters such as the magnetic field, bias potential and temperature \cite{AR01,ACH05,GL05,SE09,MC09,KAV11,MC15,FM17,OA18,OA182}.  
 In the accompanying paper \cite{YA18L}, we have checked that our approach reproduced the correct low- and high-magnetic-field dependence at the zero bias.
 To further check that our approach captures the essential feature of the transport phenomena in the two-lead Kondo model, we study the conductance behavior at finite bias and magnetic field. For a given bias $V_{0}$, we calculate the current $I(t)$ and its time-averaged steady value $\bar{I}$. We then determine the  differential conductance $G$ by calculating $\overline{I}$ for slightly modulated biases $V=V_{0}\pm\Delta V$ as follows:
\begin{equation}
G(V_{0})\simeq\frac{\bar{I}(V_{0}+\Delta V)-\bar{I}(V_{0}-\Delta V)}{2\Delta V}.
\end{equation}
We plot the typical time-evolutions of the currents relaxing to their steady values for various bias potentials $V$ in Fig.~\ref{fig_trans_p}. Figure~\ref{fig_trans_2}(a) shows the conductance behavior against magnetic field $h_z$ for zero and finite bias potentials. In the absence of bias, applying a magnetic field $h_z$ larger than the Kondo temperature $T_{\rm K}$ eventually destroys the Kondo singlet and monotonically diminishes the conductance (black circles). In contrast, with a finite bias potential $V$, it has a peak  around $h_z=V$ at which the level matching between the fermi surfaces in the two leads occurs. Importantly,  peak values of the conductance are less than the unitarity limit $2e^2/h$ since the magnetic field partially destroys the Kondo singlet as inferred from nonzero impurity magnetizations shown in Fig.~\ref{fig_trans_2}(b). 

We also plot the current-bias characteristics for different strengths of applied magnetic field in the top panel in Fig.~\ref{fig_trans_3}. As we increase the magnetic field, the current-bias curve deviates from the perfect linear characteristics (black dashed line) due to the partial breaking of the Kondo singlet. At finite bias and fixed $h_z$, the slope of the current-bias curve becomes increasingly sharper as we increase the bias $V$ as long as the magnetic field remains below the resonance $h_z<V$. Equivalently, this behavior manifests itself as the peak in the differential conductance around $h_z\sim V$  (c.f. the red curve in the bottom panel of Fig.~\ref{fig_trans_3}).  These characteristic features of the conductance under finite bias and magnetic field are consistent with  previous findings in the Anderson model  \cite{KRM02,AHKA06,JE10} and analytical results at the Toulouse point  \cite{BCJ16}. 

Several remarks are in order. Firstly, in real-space calculations  \cite{AHKA06,JE10} as performed in the present implementation, it is known that the bias potential $V$ should be kept small because of the finite bandwidth intrinsic to the lattice model.  If one chooses $V$ to be a very large value comparable to the bandwidth, the bias becomes an order of the Fermi energy and the calculations of the current and the conductance can no longer be faithful. For the parameters used in Fig.~\ref{fig_trans_3}, we find that the results faithfully converge to steady-state values as long as the bias is small enough to satisfy $V\lesssim t_{\rm h}$. Beyond that value, the current often does not reach to a steady value in the available time scale. 
Secondly,  we remark that, in the perturbative regime $V\lesssim  0.1t_{\rm h}$ in Fig.~\ref{fig_trans_3}, very small deviations of the conductance from the perfect value $G_0=2$ can be easily masked by numerical errors caused by time-dependent current fluctuations in the steady-state regime (see e.g., Fig.~\ref{fig_trans_p}). This made it somewhat difficult to test the perturbative scaling of $G$ in the limit of $V\to 0$ \cite{AR01,ACH05,SE09,MC09,KAV11,MC15,FM17,OA18,OA182}. Thirdly, as mentioned above, we found that several important features in our results are consistent with the analytical results at the Toulouse point. These include the appearance of the peak in the differential conductance around $h_z\sim V$ and the asymmetry between the conductance behavior at $V=0$ with varying $h_z$ and the one at $h_z=0$ with varying $V$. In particular, it has been argued that this asymmetry is an important feature which is absent in the ``conventional" bosonization treatment \cite{BCJ16}. Yet, there seem to be differences between the asymmetric behavior found in our results and that found in the above reference. For instance, while our results (c.f. the $h_z=0$ curve in Fig.~\ref{fig_trans_3}) indicate the negative curvature $d^2G/dV^2<0$ in the intermediate regime $V\sim T_{\rm K}\simeq 0.44t_{\rm h}$, Ref.~\cite{BCJ16} has predicted the opposite sign of the curvature with a strong linear $V$ dependence in a rather broad regime of small $V$ (c.f. the blue dashed curve in Fig. 6 in the reference). We leave the observed discrepancy between the lattice Kondo results presented here and the analytical results at the Toulouse point as an interesting open question.


\section{Conclusions and Outlook}\label{secConclusion}
Motivated by the original ideas by Tomonaga \cite{TS47} and Lee, Low and Pines \cite{LLP53}, we have developed an efficient and versatile theoretical approach for analyzing the ground-state properties and out-of-equilibrium dynamics of quantum spin-impurity systems. A key idea of this approach is to introduce a canonical transformation that decouples the impurity from the bath degrees of freedom such that the impurity dynamics is completely frozen in the transformed frame. We obtain this transformation using the conserved parity operator corresponding to the discrete symmetry of the spin-impurity Hamiltonians. 
By combining the canonical transformation with Gaussian wavefunctions for the fermionic bath, we obtain a family of variational states that can represent nontrivial correlations between the impurity spin and the bath.
In the accompanying paper \cite{YA18L}, we have presented the unitary transformation decoupling the spin-1/2 operator and the bath, and provided strong evidence that our approach correctly captures the nontrivial ground-state and out-of-equilibrium properties in spin-impurity problems. In this paper, we presented the complete details of our variational method and benchmarked its accuracy by comparing to the results obtained with the MPS approach and the exact solution via the Bethe ansatz. Furthermore, we have analyzed out-of-equilibrium dynamics of the two-lead Kondo model in further detail by calculating its long-time spatiotemporal dynamics and studying the  conductance behavior at finite bias and magnetic fields. 

Let us summarize the key advantages of the proposed theoretical approach. Firstly, this approach is versatile and can be applied to  generic spin-impurity models, including systems with long-range spin-bath interactions and disorder. The canonical transformation that we used relies only on the existence of the elemental parity symmetry, and Gaussian states can be used to describe both the fermionic and bosonic baths.
The simplicity of our variational approach can provide new physical insight into fundamental properties of challenging impurity problems.
Secondly, our method can be used to predict spatiotemporal dynamics of the total system including both impurity and bath in long-time regimes, which were challenging (if not impossible) to explore in the previous approaches \cite{SU11,RA12}. This capability allows us to reveal new types of out-of-equilibrium phenomena in SIM, e.g., non-trivial crossovers in the long-time dynamics which originate from the nonmonotonic RG flows of equilibrium systems as found in the accompanying paper \cite{YA18L}. 
Thirdly, we note remarkable efficiency of our approach, as we achieve accuracy comparable to the MPS-based method using several orders of magnitude fewer variational parameters.
This suggests that our variational states can represent nontrivial impurity-bath correlations in a very compact way. Finally, in contrast to several previous methods, our approach can be used without relying on the bosonization, which requires introducing a cut-off energy and using a strictly linear dispersion. This advantage is particularly important in view of recent developments of simulating quantum dynamics such as in ultracold atoms \cite{RL17,EM11,CM12,FuT15,KAM16,MM15,YA15,RY16,AR05,PD11,BJ13,NY13,NY16,NM152,ZR15,ZR16,KM17} and quantum dots \cite{DFS02,THE11,LC11,IZ15,MMD17}, which allow quantitative comparisons between theory and experiments on both short and long time scales.

Our variational approach can be generalized in several ways. Owing to its versatility, the proposed approach can be straightforwardly generalized to multi-channel systems  \cite{RMP07,IZ15,BZ17,MAK12,MAK16,MAK17}, disordered systems \cite{EM96}, interacting bath such as the Kondo-Hubbard models  \cite{TH97}, and long-range interacting systems  \cite{KKS17,CF17} as typified by the central spin problem \cite{WZ07} that is relevant to nano-electronic devices such as quantum dots \cite{LD98}.  Another promising directions are generalizations to bosonic systems \cite{FGM04,FS06,FFM11,FT15}, the Anderson model and two-impurity systems \cite{SP18}, which will be published elsewhere. Extending the present approach, it is also possible to calculate the frequency-domain quantities such as the spectral function \cite{AR03,WA09}.
Our approach can be also extended to study driven systems and quantum pumping. Most of the previous works in this direction were restricted to noninteracting electrons  \cite{FR09,CR11,DM15,PY16}.
Our approach will allow for studying full distribution function in charge transport; previous studies have been mainly limited to either noninteracting  models \cite{LSL96,NYV02} or one dimensional systems that allow bosonization \cite{GDB10,GDB10L}.
In this paper, we focused on the pure Gaussian state to represent coherent dynamics of an isolated system at the zero temperature. Generalizing our method to Gaussian density matrices will make it possible to explore finite temperature systems \cite{CTA00,BR01,CPH16} and, together with the variational principle for master equations  \cite{CJ15,WH15}, to study Markovian open quantum systems subject to dissipation \cite{DAJ14,BG13,LR16,TT17} or continuous measurements \cite{HC93,LTE14,YA17nc,PYS15,YA18fcs}. Extending our approach to multiple impurities  \cite{JC81,JBA87,GA99,ZZ06,SP18} would allow for studying the most challenging problems in many-body physics like competing orders in strongly correlated fermions  \cite{VM00} and confinement in lattice gauge theories  \cite{WKG74,KJ75}. 
 We hope that our work stimulates further studies in these directions. 

\begin{acknowledgements}
We acknowledge Carlos Bolech Gret, Adrian E. Feiguin, Shunsuke Furukawa, Leonid Glazman, Vladimir Gritsev, Masaya Nakagawa and Achim Rosch for fruitful discussions. Y.A. acknowledges support from the Japan Society for the Promotion of Science through Program for Leading Graduate Schools (ALPS) and Grant No.~JP16J03613, and Harvard University for hospitality.  
T.S. acknowledges the Thousand-Youth-Talent Program of China. J.I.C. is supported by the ERC QENOCOBA under the EU Horizon 2020 program (grant agreement 742102).  E.D. acknowledges support from Harvard-MIT CUA, NSF Grant No. DMR-1308435, AFOSR Quantum Simulation MURI, AFOSR grant number FA9550-16-1-0323, the Humboldt Foundation, and the Max Planck Institute for Quantum Optics.
\end{acknowledgements}

\bibliography{reference}
\end{document}